\documentclass[aps,pre,reprint,superscriptaddress,showpacs]{revtex4-1}
\usepackage[utf8]{inputenc}
\usepackage{graphicx}
\usepackage{epstopdf}
\usepackage{amsmath}
\usepackage{amssymb}
\usepackage{color} 

\DeclareMathOperator{\sgn}{sgn}

\begin{document}


\title{Chaos on the conveyor belt}

\author{Bulcs\'u S\'andor} 
  \affiliation{Babe\c{s}-Bolyai University, Dept. of Physics, RO-400084, Cluj-Napoca, Romania }
  \affiliation{E\"otv\"os Lor\'and University, Institute for Theoretical Physics, H-1117, Budapest, Hungary }

\author{Ferenc J\'arai-Szab\'o}
  \affiliation{Babe\c{s}-Bolyai University, Dept. of Physics, RO-400084, Cluj-Napoca, Romania }

\author{Tam\'as T\'el}
  \affiliation{E\"otv\"os Lor\'and University, Institute for Theoretical Physics - HAS Research Group, H-1117, Budapest, Hungary }

\author{Zolt\'an N\'eda }
  \email[]{zneda@phys.ubbcluj.ro}
  \affiliation{Babe\c{s}-Bolyai University, Dept. of Physics, RO-400084, Cluj-Napoca, Romania }

\date{\today}

\begin{abstract}

The dynamics of a spring-block train placed on a moving conveyor belt is investigated both by simple experiments and 
computer simulations. The first block is connected 
by spring to an external static point, and due to the dragging effect of the belt the blocks undergo complex stick-slip dynamics.
A qualitative agreement with the experimental results can only be achieved by taking into account the spatial inhomogeneity of the friction force on the belt's
surface, modeled as noise.
As a function of the velocity of the conveyor belt and the noise strength, the system exhibits complex, 
self-organized critical, sometimes chaotic dynamics and phase transition-like behavior. Noise induced chaos and intermittency is also observed. 
Simulations suggest that the maximum complexity of the dynamical states is achieved for a relatively small number of blocks, around five. 

\end{abstract}


\maketitle

\section{Introduction}

Spring-block type systems are successfully used for modeling various complex systems which exhibit self-organization. 
Usually, the static friction coefficients exceed the dynamical ones, in systems composed of two or more blocks connected by linear springs the coexistence of 
these friction types can lead to avalanches, nonlinear dynamics or Self-Organized Criticality (SOC) \cite{Bak1987,Bak1988,Bak1989,Carlson1989}.

Spring-block type models have broad interdisciplinary applications (for a recent review see \cite{Neda2009}), 
and prove to be very useful in describing many complex phenomena in different areas of science. For example, 
they have been applied successfully to explain elements of the Portevin-Le Chatelier effect \cite{Lebyodkin1995}, the fragmentation 
obtained in drying granular materials in contact with a frictional surface \cite{Andersen1994a,Andersen1994b,Leung2000}, to 
understand the formation of self-organized nanostructures produced by capillary effects \cite{Jarai-Szabo2005,Jarai-Szabo2011}, to 
model magnetization processes and Barkhausen noise \cite{Kovacs2005},  to describe glass fragmentation \cite{Horvat2012}, and 
even for modeling highway traffic \cite{Jarai-Szabo2011a,Jarai-Szabo2012}. 

The first members of this model-family were presented in 1967 by R. Burridge and L. Knopoff \cite{Burridge1967} to explain the empirical Guttenberg-Richter law 
of the size distribution of earthquakes \cite{Gutenberg1956}. One of their original models (called BK model here) is composed of a chain of many blocks 
connected with springs and two planes, to model the sliding of tectonic plates (for a recent review see for example \cite{Kawamura2012}).
The blocks can slide with friction on the bottom surface 
and they are all connected by springs to the upper plane which is dragged with a constant velocity. This 
model presents stick-slip dynamics, and the size distribution of the slipping events exhibits a power-law scaling.

Since the pioneering work of Carlson et al. \cite{Carlson1989,Carlson1989a}, the dynamics of the BK model has been investigated with 
several type of friction forces. In these studies both chaotic dynamics \cite{Xu1994,Ryabov1995,SousaVieira1999,Szkutnik2004,Erickson2011} 
and phase transition-like phenomena \cite{SousaVieira1993} have been reported. 
 
\begin{figure}
\includegraphics[width=8.5 cm,keepaspectratio=true]{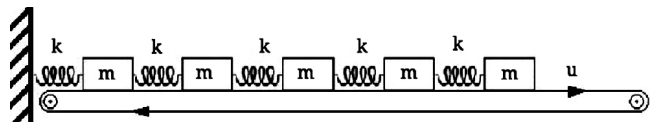}
\caption{The scheme of the investigated system. Blocks of mass $m$ connected by springs of spring constant $k$ are placed on a 
conveyor belt that is moving with a constant velocity $u$.}
\label{fig:1}
\end{figure}

The system used here has also been proposed by Burridge and Knopoff \cite{Burridge1967}, and later it was referred as "train model" \cite{SousaVieira1992}. 
As sketched in Fig. \ref{fig:1}, a spring-block chain is placed on a platform (conveyor belt) that moves with constant velocity. The first block is connected by a
spring to a static external point. As a result, due to the dragging effect of the moving platform, the blocks will undergo complex stick-slip dynamics.

The stick-slip dynamics of one block has been studied by 
analytical \cite{SousaVieira1994,Baumberger1995,Vasconcelos1996,Elmer1997}, numerical \cite{Baumberger1995,Vasconcelos1996,Elmer1997,Sakaguchi2003} 
and experimental methods \cite{Johansen1993,Baumberger1995}. 
The results for a chain composed of several blocks are contradictory from the point of view of self-organized criticality \cite{SousaVieira1992,SousaVieira1996,Vasconcelos1996,Chianca2009}. 
Undoubtedly, chaotic dynamics has been found in such systems by many authors \cite{SousaVieira1995,SousaVieira1999,Szkutnik2002,Erickson2008}. In these systems nonlinearity is introduced 
via friction forces, and several models have been studied from velocity-weakening \cite{SousaVieira1992,SousaVieira1994,SousaVieira1995,Szkutnik2002} 
to state-dependent \cite{Sakaguchi2003} friction forces. 
It has been shown that with velocity-weakening friction forces (and a constant static friction), the case of two blocks is the simplest autonomous spring-block system 
exhibiting chaos \cite{SousaVieira1995}. It was also argued, that for systems composed of many blocks, chaotic dynamics and SOC can coexist 
\cite{SousaVieira1996}.

It needs to be mentioned that a somewhat modified version of the train model has also been investigated, 
where in contrast with the former model, both the first and the last blocks are connected to static points. Additional springs are also introduced.
This system also exhibits periodic and chaotic behavior \cite{Galvanetto1998, Galvanetto2002}.

Although the proposed problem has been thoroughly investigated for one, two and many blocks, we believe that a careful experimental and 
detailed simulation study for an intermediate number of blocks can reveal new and interesting dynamical complexity.
Comparisons between analytical, numerical and experimental results exist mainly for the one block system.
An exception is the work of Burridge and Knopoff \cite{Burridge1967}, where a chain of eight blocks was thoroughly investigated. 
The main purpose of their work was, however, to investigate the distribution of the potential energy released during avalanches, and they did not considered the problem
from a dynamical systems view. Also, a general feature of previous numerical studies is that the model parameters and assumptions are arbitrarily chosen,
and usually the only argument for this is to make the calculations or the numerical code simpler.

Here we consider an experimental setup composed of $5$ blocks, and computer simulations of a simple model for systems up to 
$10$ blocks.  
In contrast to previous works, the model parameters are realistically estimated from experimental measurements.
We also incorporate the surface inhomogeneities of the conveyor belt by using a Coulomb type friction force varying stochastically about a mean.
This feature appears to be essential since the temporal intermittency observed in the experiments cannot be recovered with only deterministic friction.
A spring force with exponential cutoffs is used, to make the force profile more realistic and to avoid the collisions of the blocks with each other.

To the best of our knowledge this intermediate system size has not yet been investigated experimentally. 
The choice of intermediate system sizes is motivated by the expectation that this is the range where the theory of collective phenomena and of dynamical systems overlap
and thus tools borrowed both from statistical mechanics and chaotic dynamics become both applicable.
Despite the relatively small number of elements 
the system consists of, we surprisingly find a sharp phase transition-like behavior as a function of the conveyor belt velocity.  Interestingly, as the size of the system is increased this transition becomes smoother, a 
phenomenon that can be understood through the intermittency that was revealed in this region.  Tuning the level of disorder to a certain value, 
disorder induced phase transition-like behavior is also observed. We show that by using the Coulomb type friction forces this simple system 
exhibits a complex, chaotic dynamics accompanied by power-law type avalanche size distribution indicating SOC, noise induced intermittency 
and also noise induced chaos.

The work is structured as follows. First the used experimental setup is presented and the results are described for 
a chain of five blocks (Sec. \ref{sec:experiments}). In Sec. \ref{sec:simulation_model} the model is detailed. 
Simulation results (Sec. \ref{sec:simulation_results}) without and with a disorder in the friction force are discussed. Finally, the influence of the number 
of blocks (up to $N=10$) on the observed dynamics is investigated (Sec. \ref{sec:results_for_different_system_sizes}) and final conclusions are drawn (Sec. \ref{sec:conclusions}).  

\section{Experiments}
\label{sec:experiments}

\begin{figure}
 \includegraphics[width=6 cm,angle=0, keepaspectratio=true]{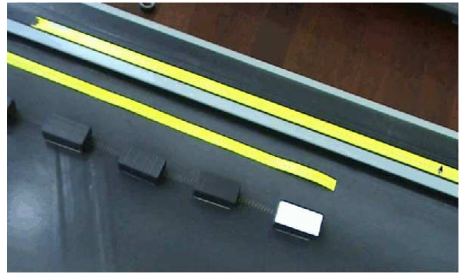}
 \caption{(color online) Snapshot of the experimental setup. The spring-block system is placed on the conveyor belt of a 
treadmill.}
 \label{fig:2}
\end{figure}

The experimental setup is sketched in Fig. \ref{fig:1}. The chain is built by $N=5$ black wooden blocks of mass 
$m=115.8$~g, and of dimensions $4$~cm$\times 8$~cm$\times4$~cm. The blocks are connected by steel springs of rest length 
$l=7$~cm and spring constant of $k=19.8$~N/m. The chain is placed on the conveyor belt of a treadmill (running machine) with 
adjustable speed. A digital camera is placed above the chain to record the dynamics of the blocks with $24$~fps (for a snapshot see Fig.
\ref{fig:2}). In order to make the last block clearly distinguishable from the others, and from the black colored 
conveyor belt, the top of this block is colored white.

\begin{figure}
 \includegraphics[width=4.5 cm,keepaspectratio=true]{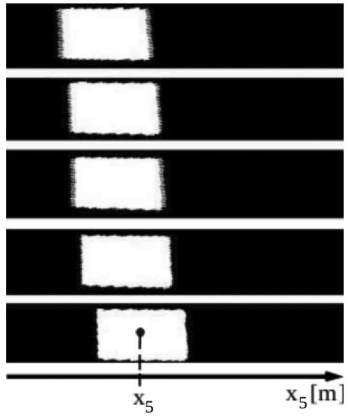}
 \caption{A processed black and white image sequence of the center of mass coordinate $x_5$ of the last ($5$th) block.}
 \label{fig:3}
\end{figure}

The video recordings are converted by a threshold operation into black and white image sequences as shown in Fig. 
\ref{fig:3}. Then, the position of the last block is detected on these images. The length unit on the image can be 
determined using the image of a tape measure, which is placed next to the chain. Accordingly, the length $x_{5}(t)$ of the chain as a 
function of time can be obtained automatically with a simple processing program.

The experiments were carried out with two values of the belt's speed: $u=0.22$~m/s and $u=0.28$~m/s. The measured average 
values of friction forces in the experiment are $F_{st_0}=1.98$~N for static friction and $F_{k_0}=0.89$~N for kinetic friction. Thus, 
the ratio of the two types of friction forces is $f_{s}=F_{k_0}/F_{st_0}= 0.45.$

\begin{figure}
 \includegraphics[width=7.8 cm,keepaspectratio=true]{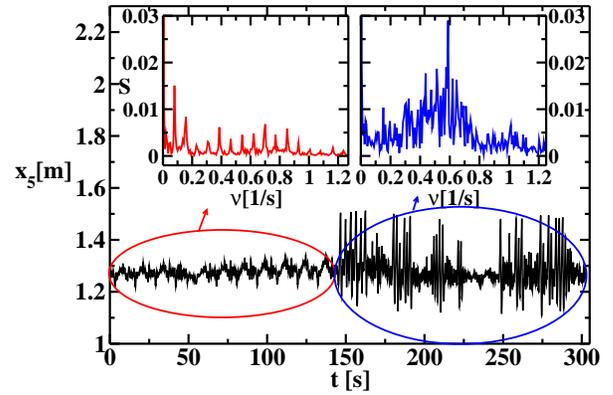}
 \caption{(color online) Experimental result for the chain length $x_{5}(t)$ as a function of time when the velocity of the 
conveyor belt is $u=0.28$~m/s. The power spectrum S of the FT is computed separately for the two clearly distinguishable regions and are
presented in the insets.}
 \label{fig:4}
\end{figure}

In both experiments the system is initialized with blocks sticking to the belt and with undeformed springs. 
Initially, the conveyor belt is at rest. After it is set in motion, length oscillations of very large amplitude occur on a time-interval up to $100$~s. 
In order to ensure that these initial transients are not considered and a kind of steady state has set in, a time-interval of $t_{\rm trans}=200$~s is 
discarded in the data processing.
The recorded data for the position of the last 
block $x_5(t)$ as a function of time is then analyzed. For the velocity $u=0.28$~m/s, the recorded data reveal (Fig. \ref{fig:4})  
two qualitatively different temporal behaviors. In the domain characterized by small amplitudes in $x_5(t)$, the 
dynamics of the system is nearly periodic, while in the domain of large amplitudes the last block exhibits a chaotic-looking behavior. 
We recall again, that the initial periodic-like behavior in this figure is not a result of the lasting transients, since before this, several chaotic-like 
regions were already present during the large amplitude bursts following the start of the conveyor belt.

The differences in the dynamics for these domains can be better understood from the Fourier 
Transforms (FT). We have thus computed separately the FT for the two well distinguishable regions (see Fig. \ref{fig:4}). 
It is clearly observable that for the small amplitude interval the power spectrum $S$ of the FT has peaks at equal distances 
suggesting periodic dynamics, while in the chaotic region the power spectrum presents a quasi-continuous distribution.

\begin{figure}
 \includegraphics[width=7.8 cm,keepaspectratio=true]{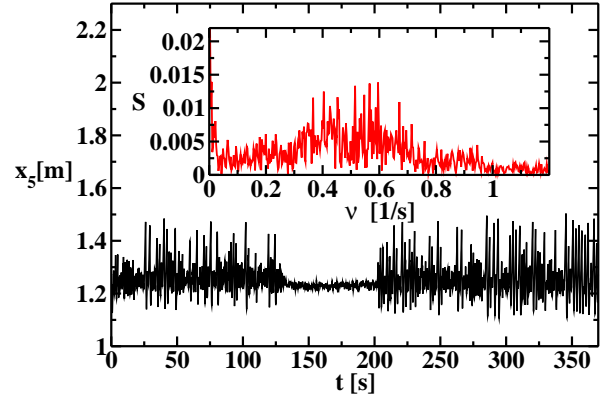}
 \caption{(color online) Experimental result for the chain length $x_{5}(t)$ with $u=0.22$~m/s. The power spectrum S of the FT for $x_5(t)$ is presented in the inset.}
 \label{fig:5}
\end{figure}

As can be seen in Fig. \ref{fig:5}, for $u=0.22$~m/s the dynamics of the system is intermittent. In 
the first 125 seconds there are large "avalanches" that result in a simultaneous movement of blocks (they slip and 
stick to the belt together). In this regime, the length of the chain evolves in time with fluctuations of large amplitudes. Then, without 
any external influence, the dynamical behavior changes abruptly, and for more than one minute the system remains in a nearly 
steady state, when all the blocks are slipping continuously relative to the belt. After this time interval, the behavior of the system looks chaotic 
again, of the type it was at the beginning of the recorded dynamics. The power-spectrum of the FT for the whole plotted interval is similar to the 
one observed for the chaotic regime for $u=0.28$~m/s.

\section{A simple model}
\label{sec:simulation_model}

The model contains the same elements (blocks and springs) as the experimental setup. The main challenge in the 
modeling effort is however the quantification of the friction and spring forces and the numerical integration of the equations of motion. 
Let us consider a chain formed by $N$ identical blocks of mass $m$. The blocks are connected by identical springs of linear spring 
constant $k$ and undeformed length $l$. As Fig. \ref{fig:1} indicates the motion of the blocks of the model is one dimensional
and the transverse oscillations present in the experiment are thus neglected.

In the model, dimensionless units are taken.
We have chosen these units so that $m=1$, $k=1$, and $l=50$. The value for $l$ was chosen for the sake of an easier 
graphical visualization (spring length corresponding to $50$ pixels). Being motivated by the experiment, we cannot avoid using a nonzero rest length for the springs.
In order to have these dimensionless values to
correspond to the experimental situation, the units are chosen as:  $[m]=0.1158$~kg for mass, $[k]=19.8$~N/m for spring 
constant, and $[l]=1.4\cdot 10^{-3}$~m for length. The units of the other quantities follow from 
dimensional considerations. The time, velocity and force units are thus, $[t]=\sqrt{[m]/[k]}=0.0765$~s, $[u]=[l]/[t]=0.0183$~m/s, and $
[F]=[k]\cdot [l]=0.0277$~N, respectively.

The equation of motion for the $i$th block of the chain is written as
\begin{equation}
\ddot{x_i}=F_e(\Delta l_{-})-F_e(\Delta l_{+})+F_f[v_{r_i},F_e(\Delta l_{-})-F_e(\Delta l_{+})],
  \label{eq:newton}
\end{equation}
where $\Delta l_{-}=x_i-x_{i-1}-50$ and $\Delta l_{+}=x_{i+1}-x_{i}-50$, respectively, and $v_{r_i}$ is the relative velocity of the block to the conveyor belt.
The elastic force, $F_e$, and the friction force, $F_f$, are defined below. 

\begin{figure}
 \includegraphics[width=7.8 cm,keepaspectratio=true]{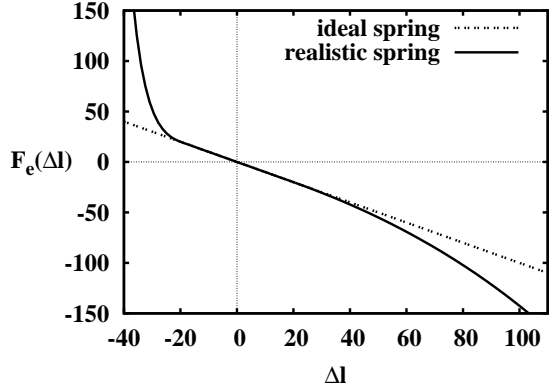}
 \caption{The used nonlinear spring force profile, in comparison with the ideal spring force profile. }
 \label{fig:6}
\end{figure}

The elastic force $F_e$ of any spring is linear, up to a certain deformation value, $\Delta l_{max}$. For higher deformations, 
this dependence is assumed to become exponential, with an exponent bigger (in modulus) for 
negative deformations (see Fig. \ref{fig:6}). Accordingly,

\begin{equation}
F_{e}(\Delta l) = 
{
\begin{cases}
  \Delta l_{max} + \frac{1}{b_1}e^{b_1(|\Delta l|-\Delta l_{max})}-\frac{1}{b_1}, \\  \hspace{3 cm} \text{if} \ \Delta l < -\Delta l_{max}, \\
  -\Delta l, 	\\ \hspace{3 cm} \text{if} \ -\Delta l_{max} \leqslant \Delta l \leqslant \Delta l_{max},\\
  -[\Delta l_{max}+\frac{1}{b_2}e^{b_2(|\Delta l|-\Delta l_{max})}-\frac{1}{b_2}], 	\\ \hspace{3 cm} \text{if} \  \Delta l > \Delta l_{max},
\end{cases}
}
\end{equation} 
where we have chosen $\Delta l_{max}=20$, $b_1=0.2$ for $\Delta l<0$ and $b_2=0.01$ for $\Delta l \geqslant 0$. With these choices 
the model is more realistic since the nonlinearity of spring forces is taken into account and collisions between blocks are 
avoided. At the same time the choice of parameters $b_1$ and $b_2$ is somewhat ad-hoc and we have not carried out a systematic change of them,
other than the fact that the average chain lengths are thus comparable with those seen in the experiment with an error less than 10\%.

According to a review of experimental results presented in Ref. \cite{Johansen1993}, "the classical friction law, where the friction force is proportional to the 
load, will only exist in an average sense" and it is important to take into account that in many cases the friction is determined by 
surface asperities, which means that "deterministic friction-velocity relations at best only exist in an average sense". In 
\cite{Johansen1993}, the authors also argue that in the case of stick-slip dynamics on a plane surface, the friction forces may have normal distributions. 

In the present paper Coulomb's law of friction \cite{Bowden1954} is used with a noisy extension reflecting the surface irregularities. 
Both the static and kinetic friction forces are independent of velocity modulus. A block remains in a stick state until the resultant external force $F_{ex}$ exceeds the value of the static friction force, $F_{st}$. For higher external force values the block starts to slide in the presence of the kinetic friction force $F_{k}$. 
We assume that, the ratio of the static and kinetic friction forces $F_{k}/F_{st}=f_s$ is constant. The friction force $F_{f}$ 
acting on the block depends both on $\sgn{(v_r)}$, where $\sgn{}$ is the signum function and $v_r$ is the block's speed relative to the conveyor belt, and on the value of the resultant external forces $F_{ex}$ acting on it. 
In our 1D setup, the friction force orientation is given only by its sign:

\begin{equation}
  F_{f}(v_r, F_{ex})=
  \begin{cases}
  -F_{ex},		& \text{if} \ v_r=0, |F_{ex}| < F_{st}, \\
  -\sgn{(v_r)}f_{s}F_{st},	& \text{if} \ v_r\neq 0,
  \end{cases}
\end{equation} 
where $v_r=v-u$  and $v$ is the velocity of the block relative to the laboratory frame.

Since the surface of the conveyor belt is not perfectly smooth, the already mentioned argument of Ref. \cite{Johansen1993} 
is implemented by using randomly distributed friction forces. 
This is incorporated into the model by randomly generating a new static friction force value for every different position of the block relative to the belt.
These random friction force values are generated according to a normal distribution with a fixed mean $F_{st_{0}}$ and standard deviation: $\sigma$.  
This standard deviation (together with the integration time-step, which is however the same for all simulations), will quantify the amount of disorder introduced in the model.
Although the origin of the disorder is the spatial inhomogeneity of the belt's surface, from the point of view of the dynamics the fluctuations of the friction force
appear as temporal noise. We shall thus call $\sigma$ as the noise strength.

\begin{figure}
 \includegraphics[width=7.8 cm,keepaspectratio=true]{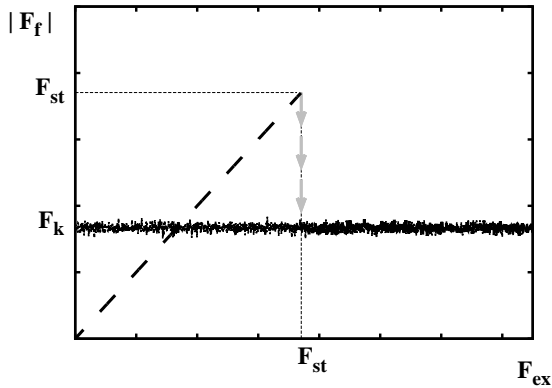}
 \caption{The stochastic version of the Coulomb friction model for one stick-slip event.  Thick dashed line represents the situation when the 
block is stuck to the belt (the resultant external force $F_{ex}$ is smaller than the maximal static friction force value $F_{st}$). 
Continuous line corresponds to the situation when the block is sliding and indicates the fluctuation of the kinetic friction force $F_k=f_sF_{st}$ during 
such an event. Dashed gray arrows indicate the sudden change of the friction force, when the block starts to slip.}
 \label{fig:7}
\end{figure}

The kinetic friction force value is always 
automatically updated together with the static friction force value, using their fixed ratio $f_s$. 
As a result, both the kinetic and static friction 
forces will fluctuate in time during the sliding of the particular block. In Fig. \ref{fig:7} we illustrate the variation of the 
friction force  as a function of the  resultant external force $F_{ex}$ for one stick-slip event. 
Such type of friction forces also proved to be useful for modeling highway traffic \cite{Neda2009,Jarai-Szabo2011a}.

In order to use the same friction force value as in the experiments, in the dimensionless units the average value of the static friction 
force is $\langle F_{st} \rangle=F_{{st}_0}=71.4$. The ratio of the two types of friction forces is also chosen in agreement with 
the experiments as $f_s=0.45$.

Computer simulations of the model start from initial conditions similar to the experimental ones. The blocks are placed on the 
belt with undeformed springs between them. At this initial setup the blocks are stuck to the belt which means that their initial 
velocity in the laboratory reference frame is equal to the conveyor belt's constant velocity $u$. Computer graphics is also used to 
visualize in real-time the dynamics of the blocks. 

The numerical methods used to solve Newton equations (\ref{eq:newton}) are presented in the Appendix. The time-step of integration is fixed to
$dt=0.001$.

In order to characterize statistically the dynamics of the chain, a parameter measuring the fluctuation of the chain length 
in dynamical equilibrium is introduced. The chain's length is determined by the position of the last block in the row. 
The {\em disorder parameter} $r$ is defined as the standard deviation of the coordinate ${x_N}$
compared to the time averaged length $\langle x_{N} \rangle$ of the chain: 

\begin{equation}
 \label{eq:order_parameter}
 r=\frac{\sqrt{\langle x_{N}^2 \rangle - {\langle x_{N} \rangle}^2}}{\langle x_{N} \rangle}.
\end{equation}

This quantity takes on large values for large fluctuations which explains the term disorder parameter.
Among other relevant dynamical measures, this disorder parameter is investigated as a function of the belt velocity $u$ (Section 
\ref{sec:results_without_noise}.), the noise level $\sigma$ (Section \ref{sec:results_with_noise}.), and the number of blocks $N$ 
(Section \ref{sec:results_for_different_system_sizes}.).

\section{Simulation results}
\label{sec:simulation_results}

\subsection{Results without noise}
\label{sec:results_without_noise}

\begin{figure}
  \includegraphics[width=7.5 cm,keepaspectratio=true]{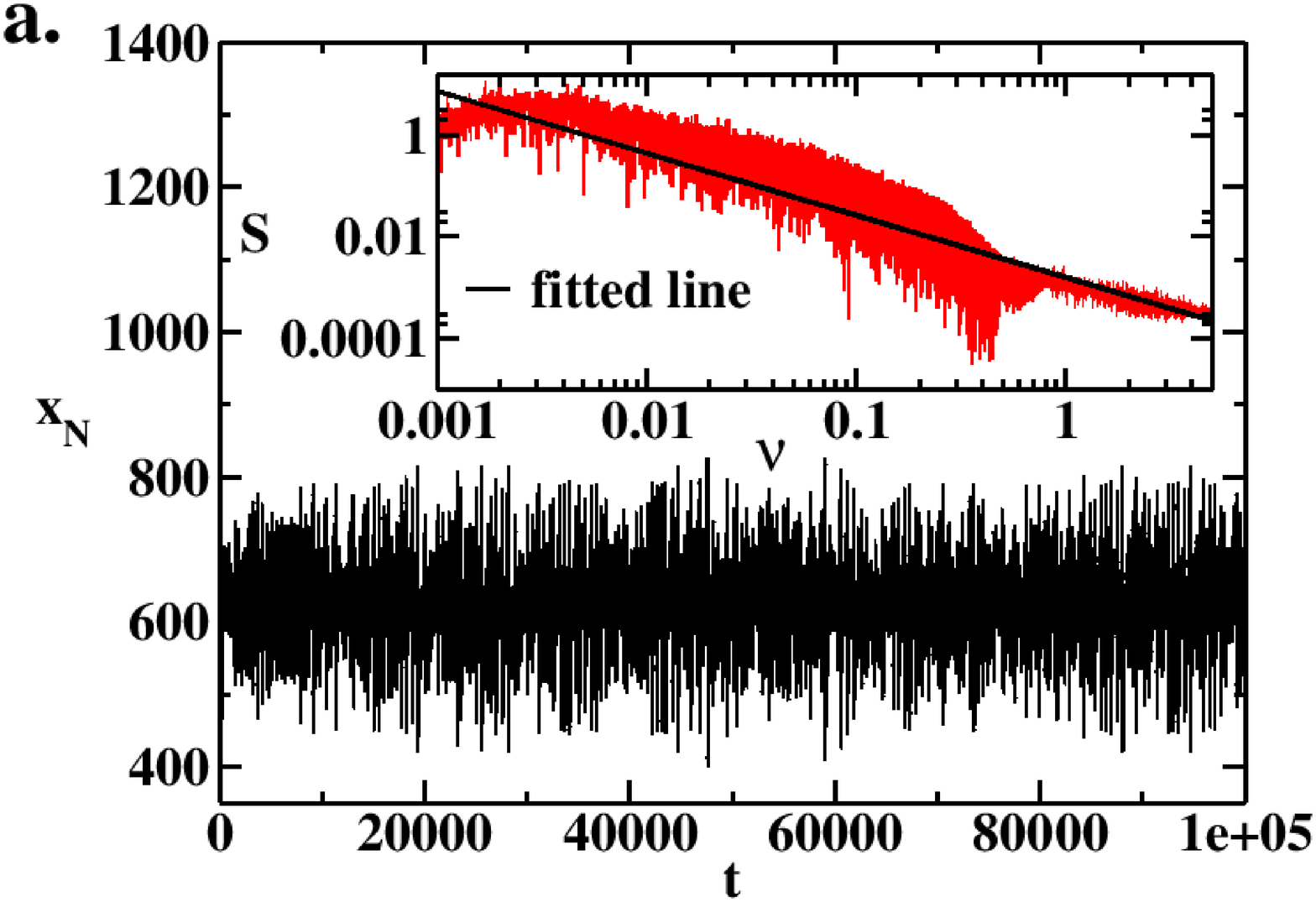}
  \includegraphics[width=7.5 cm,keepaspectratio=true]{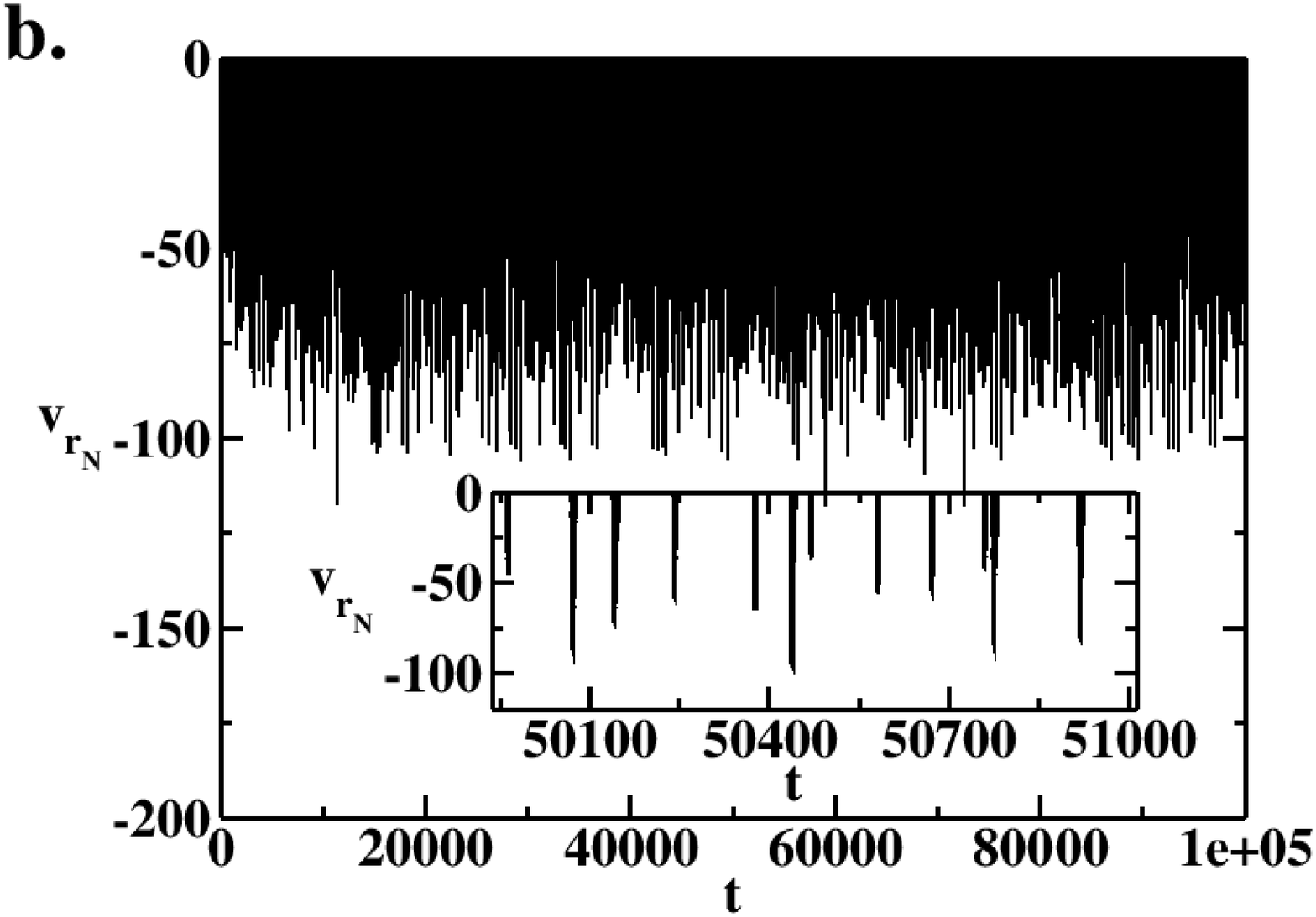}
 \caption{(color online) Computer simulation results for $u=1$, $\sigma=0$, and $N=5$.
 a. The position of the last block $x_N$  as a function of time. Inset: the power spectrum of the 
 corresponding FT. The fitted line corresponds to a power law: $S(\nu)=0.0016\cdot \nu^{-1.21}$.
 b. The relative velocity $v_{r_N}$  of the last block as a function of time. 
 The inset presents a zoom of a shorter time interval. }
 \label{fig:8}
\end{figure}

In the deterministic case, as the conveyor belt is started, the whole system moves together with the belt until the first 
block starts to slip. This slipping moment is determined simply by the value of the static friction force $F_{st}$ and the belt velocity $u
$. The block sticks again to the belt when its relative velocity $v_r$ becomes zero. After that the process starts 
again. For small belt velocities this kind of behavior characterizes all the blocks in a self-organized manner. Specifically, the 
blocks are slipping together and produce "avalanches" of different sizes. Therefore, the length of the chain, defined by the position of 
the last block $x_N$ fluctuates as a function of time with largely varying amplitudes, as shown in Fig. 
\ref{fig:8}.a. 

Furthermore, as shown in the inset of Fig. \ref{fig:8}.a. the Fourier transform of the time series of $x_N$ exhibits a nearly power-law behavior with an exponent $-1.21$, 
suggesting a $1/f$ type stochastic behavior in the long time dynamics and a critically self-organized state. 
The relative velocity of the last block $v_{r_N}$ is zero, when the 
block is stuck to the belt. As the chain starts to slip, this relative velocity is rapidly increasing in absolute value as indicated in Fig. \ref{fig:8}.b and it
shows large fluctuations according to the inset of Fig. \ref{fig:8}.b.

\begin{figure}
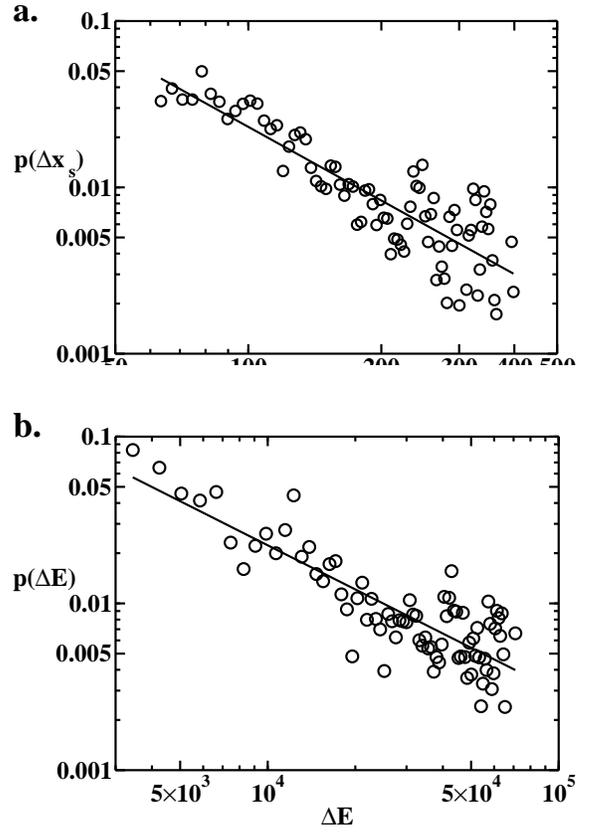

 \includegraphics[width=7.5 cm,keepaspectratio=true]{fig9a}
 \includegraphics[width=7.5 cm,keepaspectratio=true]{fig9b}
 \caption{ a. Probability density function for the size, $\Delta x_{s}$, of the avalanches.
 A power-law fit with an exponent $\alpha=-1.47$, suggests a SOC-like behavior.
 b. Probability density function for the energy $\Delta E$ dissipated during the avalanches.
  A power-law fit with  exponent $\alpha=-0.87$ is shown. 
 The simulation parameters are $u=1$, $\sigma=0$, and $N=5$.}
 \label{fig:9}
\end{figure}

For a quantitative measure of avalanches, the slip size distribution for the last block $N=5$ is computed and shown in Fig. 
\ref{fig:9}.a. The slip size $\Delta x_s$ is defined as the difference of the coordinates 
where the block starts to slip and when it stops relative to the belt. The obtained power law behavior confirms 
the presence of SOC, which appears here together with a chaotic dynamics.
The distribution of the energies dissipated during avalanches is plotted in Fig. \ref{fig:9}.b. 
The results indicates again a scaling (a sign for SOC-like behavior), although the exponent $\alpha=-0.87$ is different from the one
obtained in the seminal work of Burridge and Knopoff \cite{Burridge1967}. This difference is a natural consequence of the different nature and size of the problems.

\begin{figure}
 \includegraphics[width=8.7 cm,keepaspectratio=true]{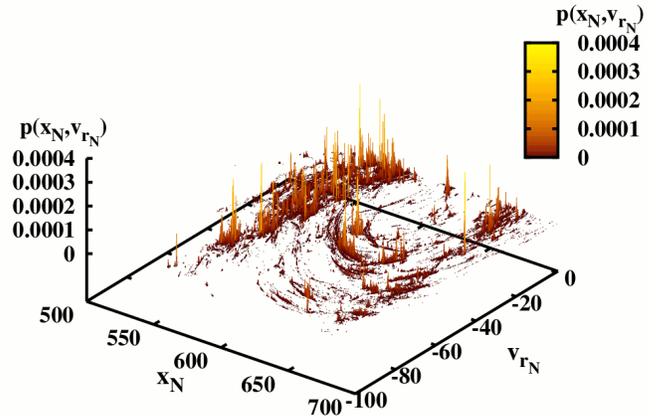}
 \caption{(color online) Natural distribution on the Poincaré section in the phase plane of the last block ($u=1$, $\sigma=0$, $N=5$). 
 The Poincaré section is taken when the  $x_{2}=304.3$ plane is crossed from right to left. The plot is made only for $v_{r_N} < 0 $ (the stuck states  with $v_{r_N}=0$ 
 are not considered for a better visibility).}
 \label{fig:10}
\end{figure}

The existence of chaos is illustrated by Fig. \ref{fig:10}, where the 
natural distribution on a Poincaré section is presented. This Poincaré section is projected onto the phase plane ($x_N, v_{r_N}$) of the 
last block, and obtained as the intersections of the system's trajectory in the phase space with the plane defined by $x_2=304.3$, 
considering only uni-directional crossings from right to left ($\dot{x}_2<0$). For a better view, 
the high probability states along the line $v_{r_N}=0$ (where the block is stuck to the belt) are not shown.
In the full system of $10$ degrees of freedom the Poincaré section is $9$-dimensional. Fig. \ref{fig:10} shows a projection of this on a plane.
It is surprising that the natural distribution on this plane is similar in appearance (both in the shape of the support and of the rather irregular distribution)
to that of the chaotic attractor of a driven one-dimensional system.
This indicates that the chaotic attractor of the full chain is rather low-dimensional.

\begin{figure}
  \includegraphics[width=7.5 cm,keepaspectratio=true]{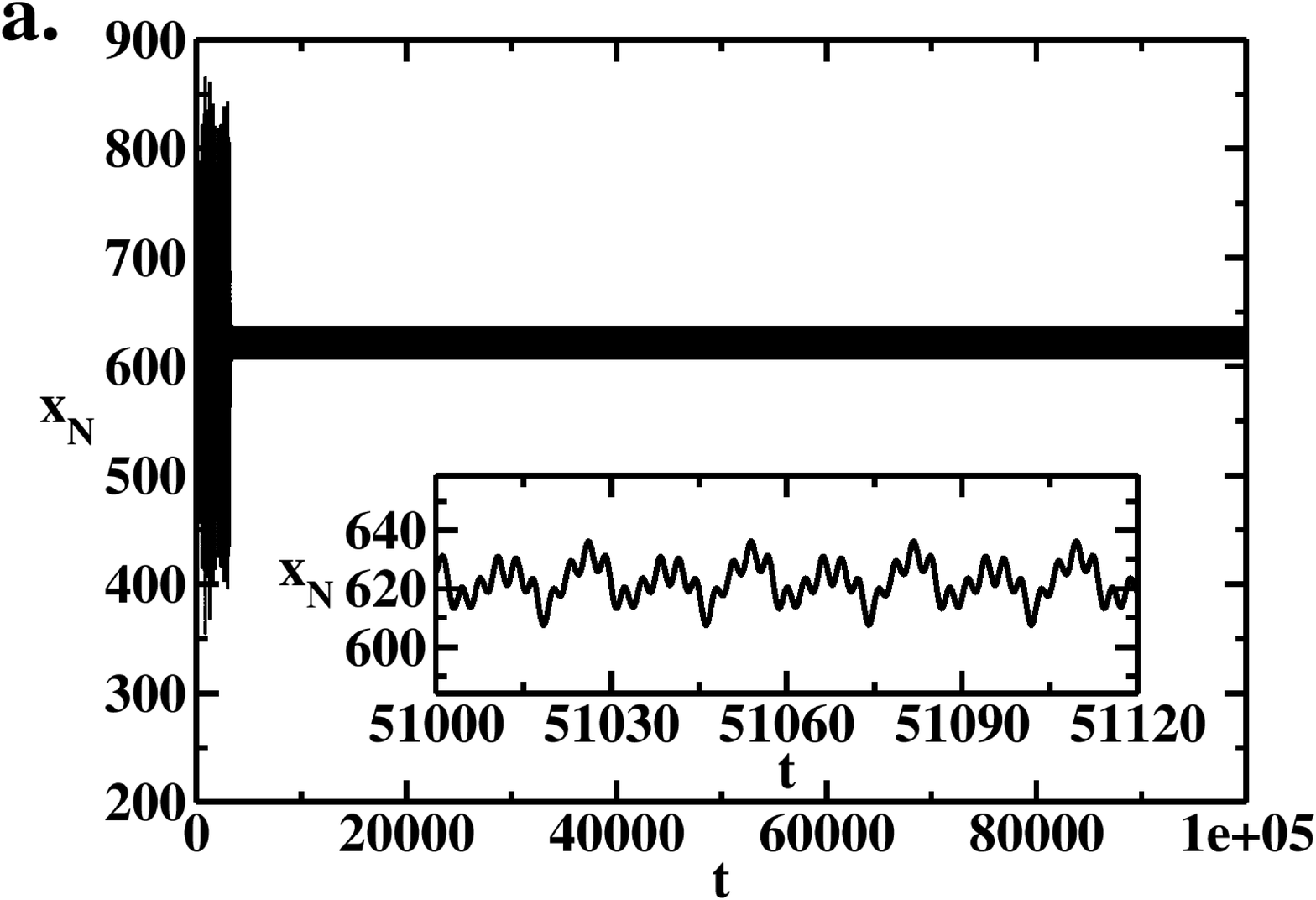}
  \includegraphics[width=7.5 cm,keepaspectratio=true]{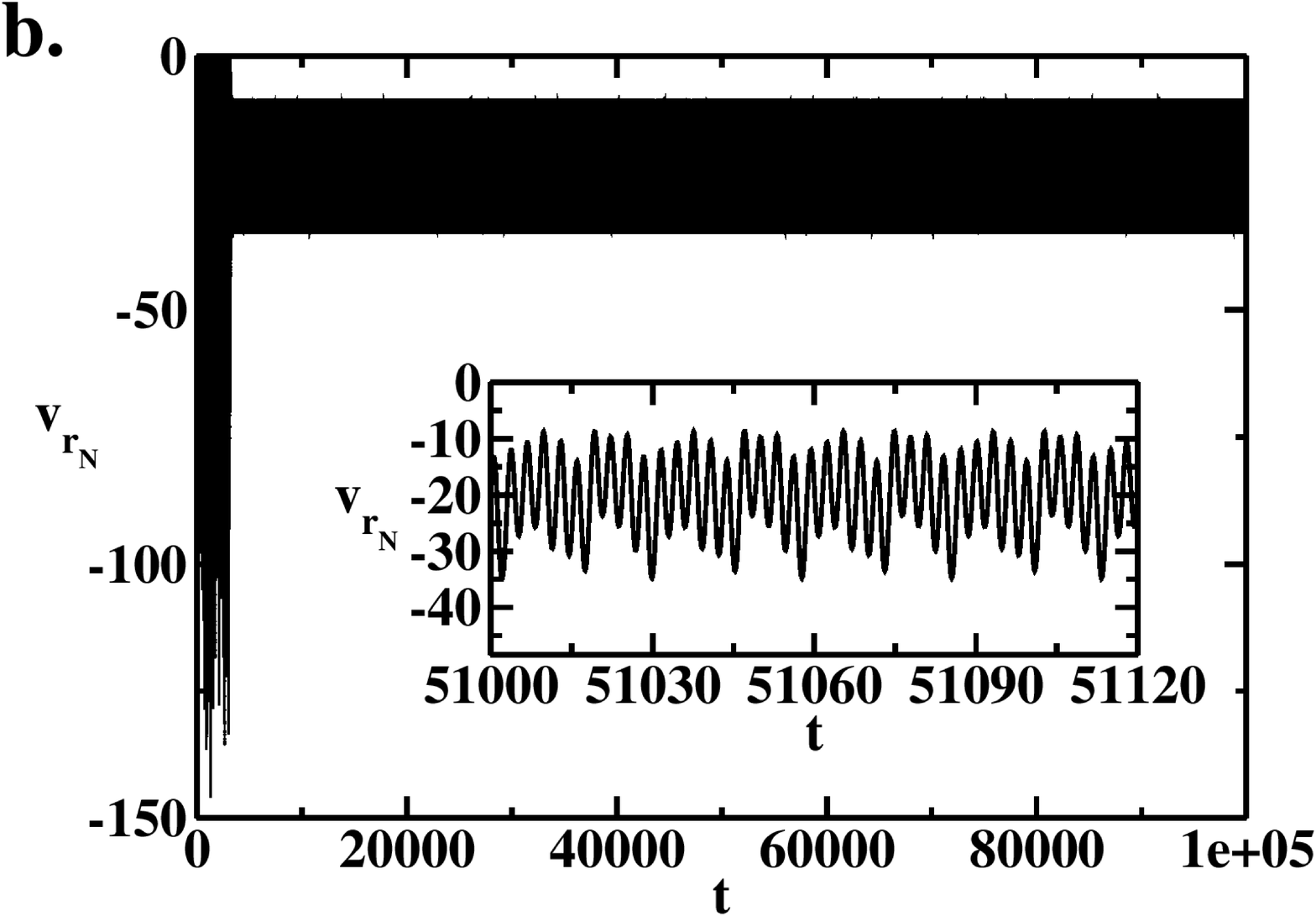}
 \caption{a. The position of the last block as a function of time for $u=20$. The inset presents a zoom on the indicated time-interval.
 b. The relative velocity $v_{r_N}$ of the last block. The last block exhibits a chaotic transient 
dynamics up to about $t=4000$, followed
by a periodic behavior with continuous slip. The inset presents a zoom on the indicated time-interval.
 The other parameters are $N=5$ and $\sigma=0$. }
 \label{fig:11}
\end{figure}

For higher values of the belt velocity, there are other possible scenarios as well. For instance, for $u=7$ the system exhibits an asymptotically 
periodic stick-slip dynamics, but for $u=15$  the behavior is aperiodic again. For the velocity interval $18\leq u\leq 50$ permanent 
chaotic dynamics never occurs for the investigated parameters (see e.g. Fig. \ref{fig:11} 
for $u=20$) and the last block exhibits a continuous slip dynamics after a certain transient time.
It can be seen in this figure that the system undergoes a transient chaotic behavior with a stick-slip dynamics,
before reaching a periodic attractor. 

\begin{figure}
 \includegraphics[width=7.5 cm,keepaspectratio=true]{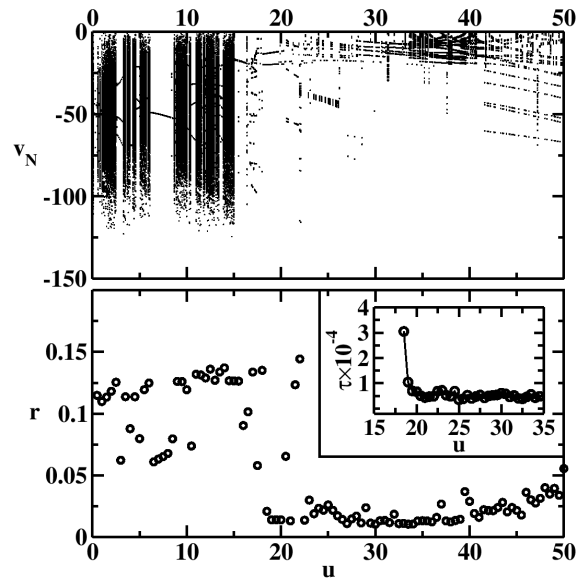}
 \caption{Top panel presents the bifurcation diagram for the velocity $v_N$ of the last block captured when $x_N=615$ and 
$\dot{x}_N<0$,
 as a function of $u$. Bottom panel illustrates the disorder parameter $r$ as a function of $u$ ($\sigma=0$ and $N=5$).
 The inset presents the average lifetime $\tau$ of transients for the $18\leq u\leq 35$ interval. 
 The missing $\tau$ values in the interval $u<18$ indicate that the lifetime of the transients is infinite, chaos is permanent. } 
 \label{fig:12}
\end{figure}

The bifurcation diagram plotted in the top panel of Fig. \ref{fig:12} 
summarizes all the cases described above. Here, the velocity of the last block $v_N$ is plotted 
as a function of the driving velocity $u$ in instances when the block goes through a fixed ($x_N=615$) position from the right.
For every value of the control parameter $u$ the 
simulation is restarted from the declared initial position.
To construct this plot, the computer simulations were run up to $t=2\times10^6$ time units, discarding a transient time of $t_{\rm trans}=10^6$.

Interestingly, the disorder parameter (\ref{eq:order_parameter}) allows for a difference to be made between these different dynamical behaviors.
Results for $r(u)$ are plotted on the bottom panel of Fig. \ref{fig:12}. 
As expected, for asymptotic chaos the disorder parameter has a high value, and in the case of 
the periodic or quasi-periodic dynamics, it is much smaller. 
In the transition region ($15< u< 18$) the value of $r$ falls from about $0.13$ to $0.02$. 
This jump of nearly one order of magnitude indicates a relatively sharp dynamical phase transition-like behavior. 
The critical speed $u_c$, defined as the midpoint of the transition region, which the phase transition-like behavior can be associated with, is $u_c=16.5$. 
(Note that in the interval $20<u<22$ there are three outlier points in the lower panel of Fig. \ref{fig:12}.
These belong however to periodic attractors of large amplitudes rather than to chaotic cases.)

In the inset of Figure \ref{fig:12},  the average lifetime $\tau$ of the chaotic transients 
preceding the periodic behavior is presented. This is measured using the method described in Ref \cite{Tel2008}. Looking at the 
simulation data, it can be seen that in the phase transition region the lifetime of transients grows considerable,
a kind of critical slowing down can be observed.
It can thus be concluded, that for $u\leqslant 15$ (except for the periodic windows) the chaotic dynamics is permanent, while for $u>18$ it has a 
transient character. These transients were neglected in computing the disorder parameter, since the discarded transient time 
($t_{trans}=10^6$) is larger by more than two orders of magnitude than the average lifetime $\tau$ of the transients.
 
\subsection{Results in the presence of noise}
\label{sec:results_with_noise}

As described earlier, the two friction forces 
are linked by the relation $F_k=f_s F_{st}$, and the values of $F_{st}$ are randomly updated for each new 
position of the blocks on the conveyor belt. The distribution of $F_{st}$ has a standard deviation $\sigma$ and a mean value $F_{st_{0}}$. 

\begin{figure}
 \includegraphics[width=7.5 cm,keepaspectratio=true]{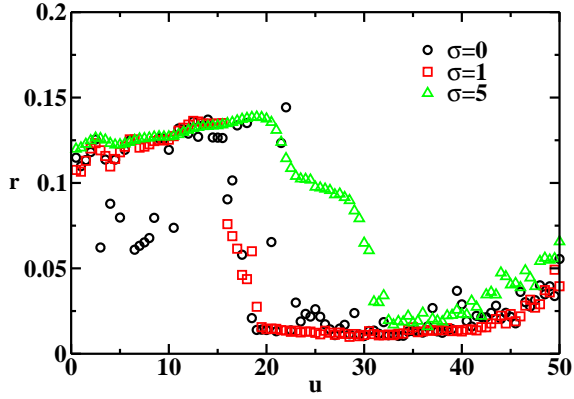}
 \caption{(color online) The disorder parameter $r$ as a function of the conveyor belt's velocity $u$ for different noise 
levels $\sigma$.}
 \label{fig:13}
\end{figure}

\begin{figure}
 \includegraphics[width=8.5 cm]{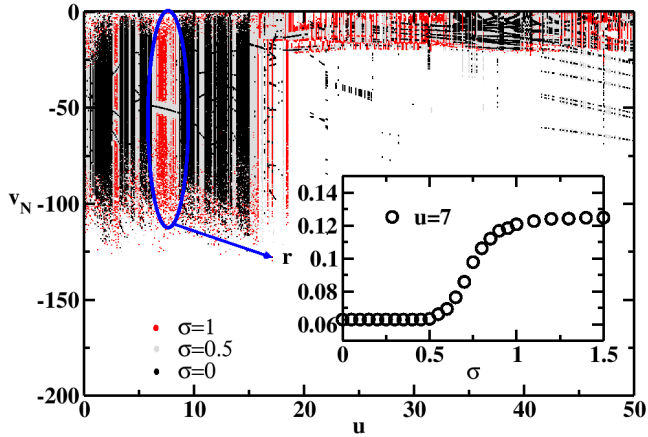}
 \caption{(color online) The bifurcation diagram, as defined in Fig. \ref{fig:12}, for different noise levels. 
 The inset presents the disorder parameter as a function of $\sigma$ for $u=7$ (a periodic window for $\sigma=0$).}
 \label{fig:14}
\end{figure}

First, a relatively low level of noise $\sigma=1$ is considered (see Fig. \ref{fig:7}). In this case, the 
phase transition-like behavior remains almost unchanged. This is clearly visible in the $r(u)$ plots of Fig. \ref{fig:13}. 
The periodic windows present for $\sigma=0$  (top panel of Fig. \ref{fig:12}) 
disappear, however, and consequently the disorder parameter has less fluctuations in the chaotic 
regime. The noisy bifurcation diagrams of Fig. \ref{fig:14} also illustrate the disappearance of the 
periodic windows as $\sigma$ increases. 

For belt velocity $u=7$ and small noise levels, the asymptotic periodic dynamics is reached after a 
transient chaotic behavior. This is possible if a non-attracting chaotic set (chaotic saddle) and periodic attractors coexist (see for example \cite{Tel2006, *Lai2011}).
Beyond a critical noise level $\sigma_{c}$, the chaotic transients turn into a permanent chaotic dynamics. This is called 
{\em noise induced chaos} \cite{IANSITI1985,HERZEL1987,BULSARA1990,HAMM1994,Tel2008}. 
Based on the results plotted in the inset of Fig. \ref{fig:14} 
the critical noise strength $\sigma _{c}$ necessary to obtain noise induced chaos is estimated as $\sigma_{c}= 0.75$.

By increasing the noise level $\sigma$ in the friction force, the critical speed $u_c(\sigma)$, for which a phase transition-like 
behavior occurs, increases with $\sigma$ as suggested by Fig. \ref{fig:13}.

\begin{figure}
 \includegraphics[width=7.5 cm,keepaspectratio=true]{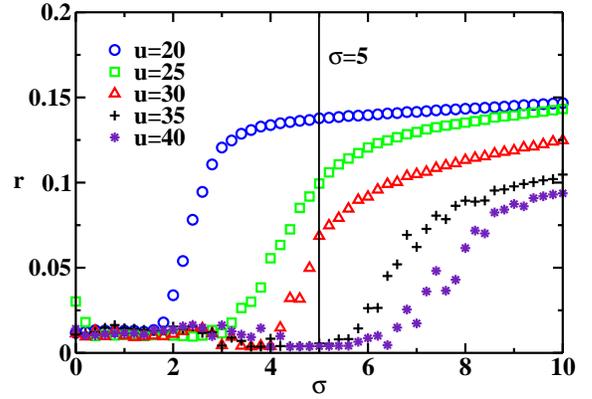}
 \caption{(color online) The disorder parameter $r$ as a function of $\sigma$ for different values of $u\geqslant20$. 
 The $\sigma=5$ line is only to guide the eye.}
 \label{fig:15}
\end{figure}

Turning the problem around, in Fig. \ref{fig:15} we plot the disorder parameter $r$ as a function of the noise level $\sigma$
for intermediate velocities $u\geqslant 20$. 
In the whole range a noise induced phase transition-like behavior emerges. Moving away from $u=20$, the sharp transition 
becomes more smooth and the transition point in $\sigma$ is shifted toward higher noise levels. A possible explanation of this effect is that 
for higher velocities a higher noise level is needed to kick the system out from the basin of attraction of the periodic attractor.

\begin{figure}
 \includegraphics[width=8 cm,keepaspectratio=true]{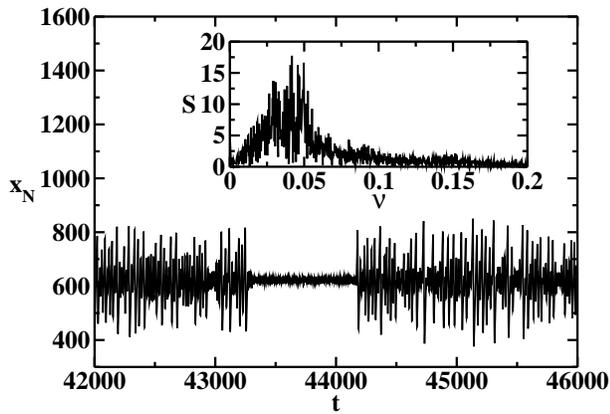}
 \caption{ The position of the last block as a function of time exhibits a similar intermittent behavior as found in the 
experiments. The inset shows the power-spectrum of the FT.  $N=5$, $u=17.5$, and  $\sigma=2.2$.}
 \label{fig:16}
\end{figure}

The model is able to reproduce the intermittent behavior observed in the experiment and presented in Fig. \ref{fig:4}. 
For example, considering $\sigma=2.2$, and belt velocities in the phase transition-like region, the system 
exhibits an intermittent dynamics (Fig. \ref{fig:16}).
Since such intermittency is not present without noise, this phenomena is called {\em noise induced intermittency} \cite{Tel2006,*Lai2011} in the chaos literature.
This can happen again only if a chaotic saddle exists in the deterministic system \cite{Tel2006,*Lai2011}.

We note that the critical speed $u_c(\sigma=2.2)$ is found to be $19$ which corresponds in dimensional units to $u_c=0.35$~m/s. 
Experimentally we have found intermittency at $u=0.22$~m/s and $u=0.28$~m/s which should belong to velocities close to the critical value. 
In view of the simplicity of the model this order of magnitude agreement is satisfactory.

The disorder parameter can be computed also for the experimental time series of $x_{5}(t)$ of Figs. \ref{fig:4} and \ref{fig:5}.
The results are: $r=0.072$ and $r=0.052$ for $u=0.22$~m/s and $u=0.28$~m/s, respectively. 
The values corresponding in the model to these dimensional velocities are:
$r=0.132$ and $r=0.134$, respectively, again in an order of magnitude agreement with the experiments.

\begin{figure*}[t]
\includegraphics[width=8.1 cm,keepaspectratio=true]{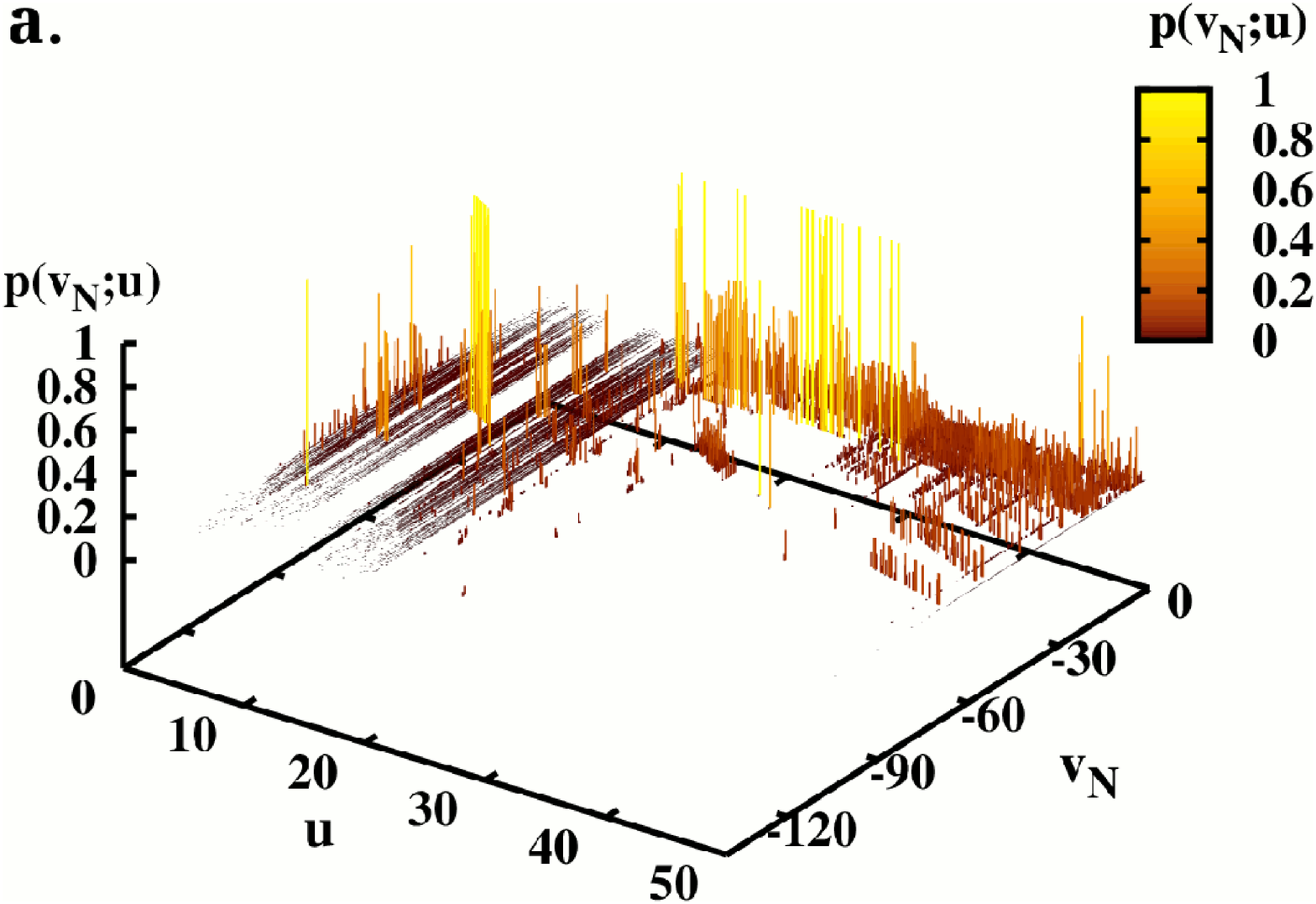}\hspace{5mm}
\includegraphics[width=8.1 cm,keepaspectratio=true]{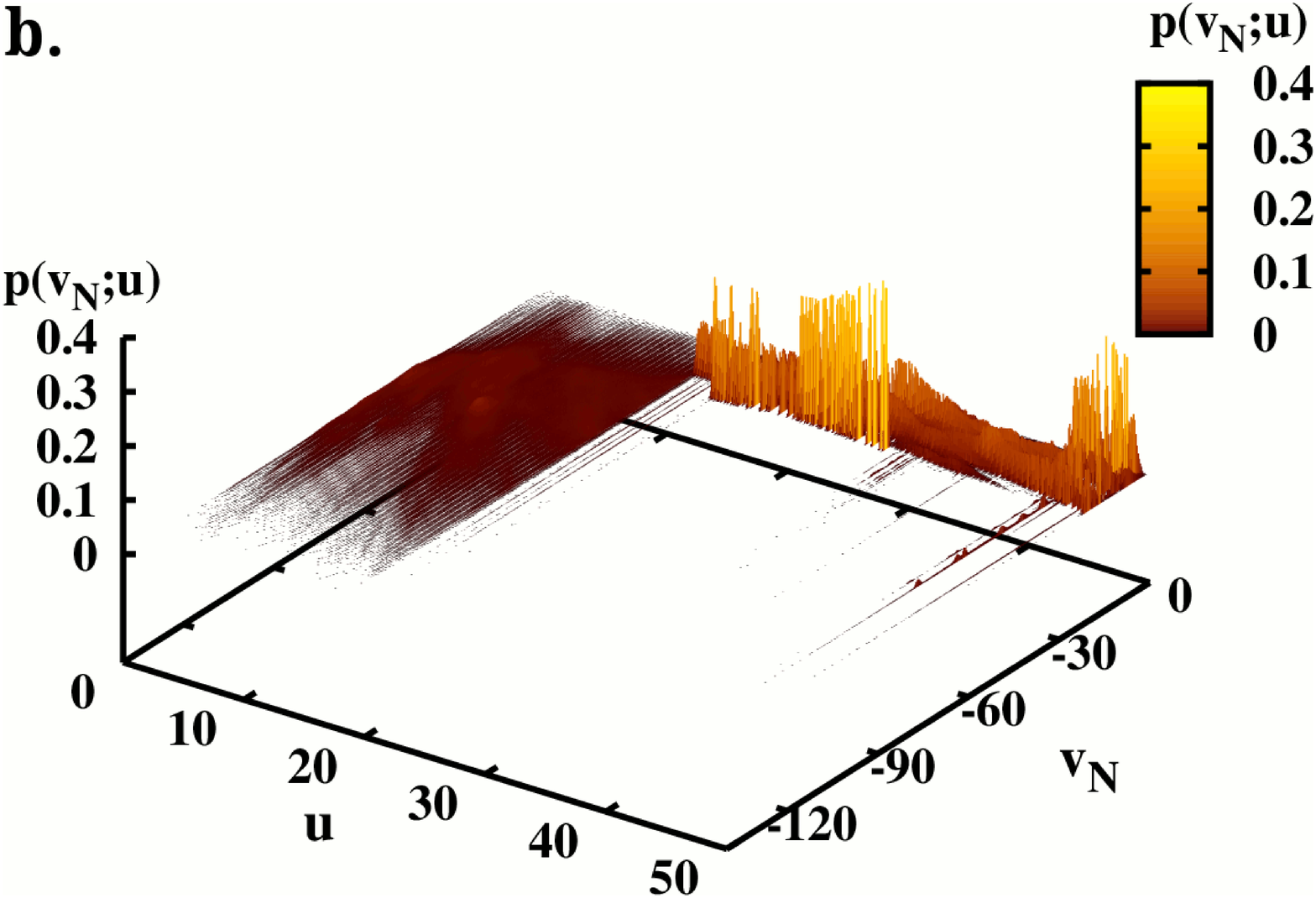}
\includegraphics[width=8.1 cm,keepaspectratio=true]{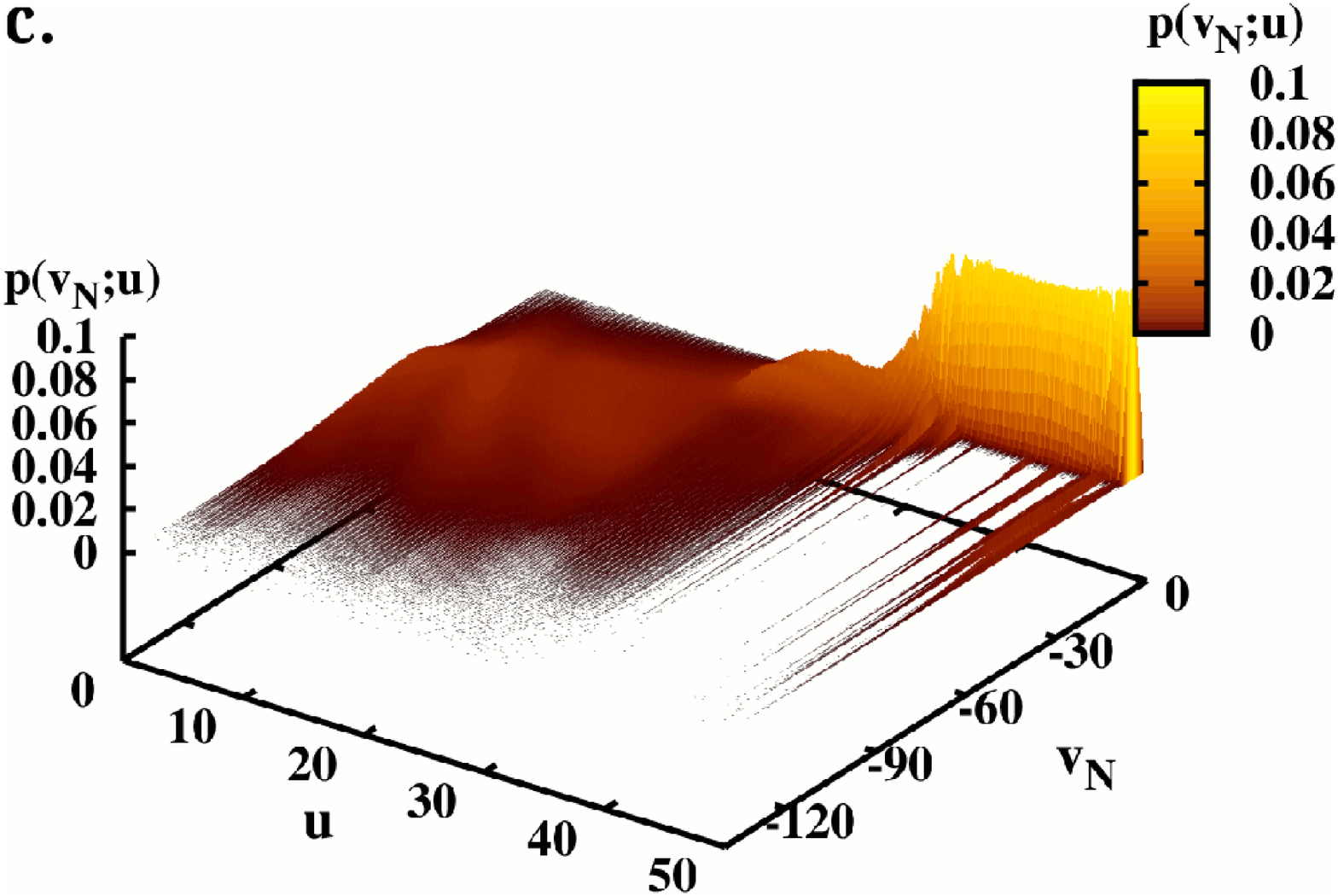}\hspace{5mm}
\includegraphics[width=8.1 cm,keepaspectratio=true]{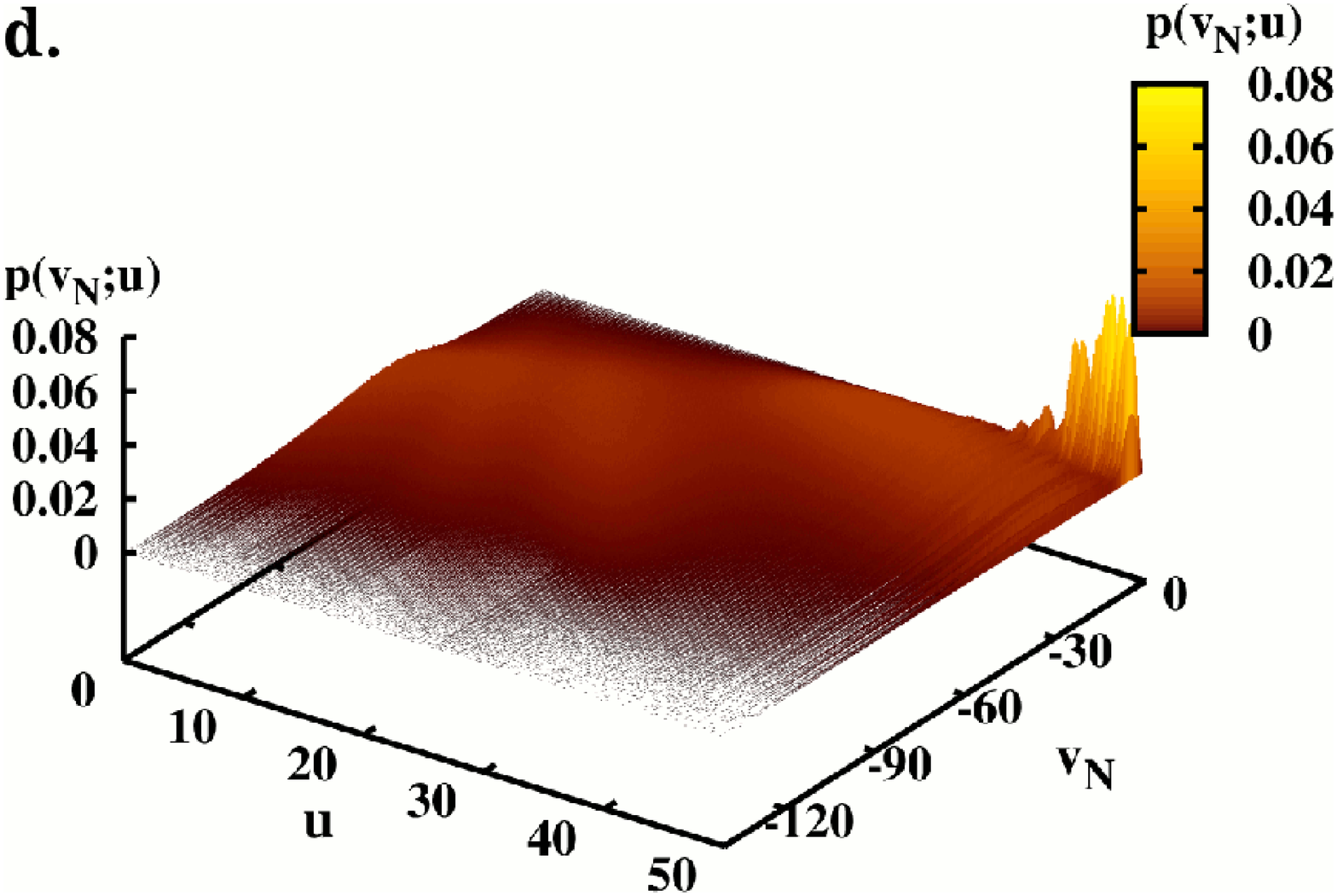}
 \caption{(color online) The natural distribution on the plane defined by $u$ and $v_N$ for increasing noise levels (N=5). 
 a. $\sigma=0$; b. $\sigma=1$; c. $\sigma=5$; d. $\sigma=10$.}
 \label{fig:17}
\end{figure*}

Another facet of the effect of noise is shown in Fig. \ref{fig:17}, where the natural distribution is projected to the 
bifurcation diagram defined in Sec. \ref{sec:results_without_noise}. The consecutive graphs are results for increasing noise levels. 
From these plots we learn that for $\sigma=1$ all the periodic windows in the $0<u\leqslant15$ interval disappear, but the phase transition
point hardly changes. As $\sigma$ is increased further, the sharp phase transition-like behavior transforms into a 
smoother one (see Fig. \ref{fig:17}.c). For $\sigma=10$ the dynamics of the system is dominated by noise (Fig. \ref{fig:17}.d).

\begin{figure}
 \centering
 \includegraphics[width=9.2cm,keepaspectratio=true]{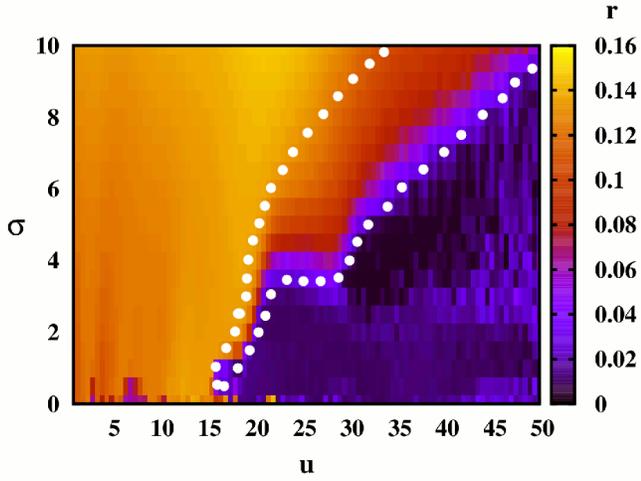}
 \caption{(color online) Disorder parameter $r$ over the parameter plane $(u, \sigma)$ for $N=5$ blocks illustrated with color-coding. 
 White dots indicate the boundaries of the parameter range where noise induced intermittency can occur.}
 \label{fig:18}
\end{figure}

The results regarding the phase transition-like behavior are summarized in Fig. \ref{fig:18} by a detailed map of the parameter space $(u ,\sigma)$.
Figures \ref{fig:13} and \ref{fig:15} correspond to sections of this map along the horizontal and vertical axes, respectively.
 The range where the system exhibits noise induced intermittency is enclosed by white dots.
 The fact that this range does not reach the $\sigma=0$ line shows, that intermittency cannot be recovered with purely deterministic friction forces.
 As the level of noise is increased, this region becomes abruptly wider for $\sigma> 4$.
 This can be interpreted as the critical noise strength below which the phase transition-like region remains sharp, and if this is exceeded, the transition becomes smoother.
 In the view of all these observations we conclude that the noise strength corresponding to the experiments is about $\sigma=1-2$.

\section{Finite size effects}
\label{sec:results_for_different_system_sizes}

\begin{figure}
 \includegraphics[width=7.5 cm]{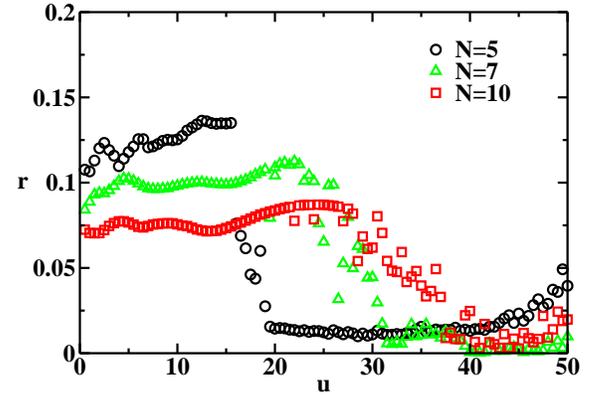}
 \caption{(color online) The disorder parameter $r$ as a function of the belt velocity $u$ for different systems sizes $N$ with the $\sigma=1$. 
 The ``transition regions'' corresponding to different system sizes are: $15< u<18$ for $N=5$,
 $23< u< 32$ for $N=7$, and $27 < u < 40$ for $N=10$.}
 \label{fig:19}
\end{figure}

We can now briefly discuss the influence of the system size $N$ on the observed phase transition-like phenomena. 
We know from statistical physics that the observed phase transitions become sharper as the system size is increased. Although the 
investigated system is a non-equilibrium dynamical system and not a system in thermal equilibrium, one might 
expect that the phase transition-like behavior sharpens for larger system sizes. In our system, 
however, the opposite happens. As shown in Fig. \ref{fig:19} for increasing system sizes, 
the sharp phase transition-like behavior is transformed into a smoother and smoother one, the transitional region broadens with $N$.

\begin{figure}
 \includegraphics[width=8.6 cm,keepaspectratio=true]{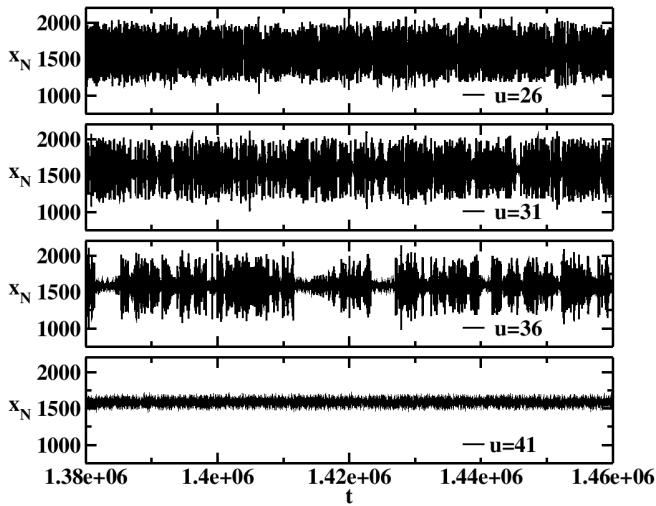}
 \caption{The position of the last block for a chain size of $N=10$,  with different velocities $u$.
 The presented time interval is after the discarded transient time  $t_{\rm trans}=10^6$ ($\sigma=0$).}
 \label{fig:20}
\end{figure}

\begin{figure}
 \includegraphics[width=7.5 cm,keepaspectratio=true]{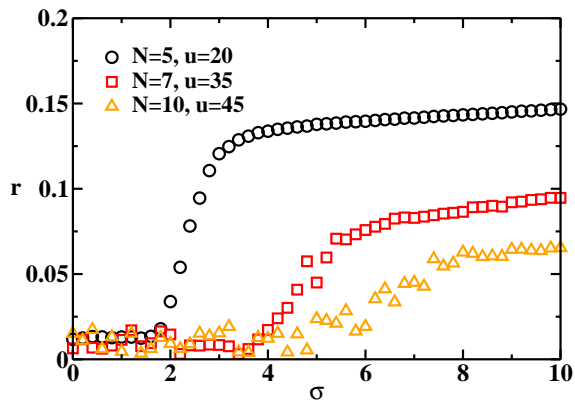}
 \caption{(color online) The disorder parameter $r$ as a function of the noise level $\sigma$ for different system sizes $N$.
 The driving velocity is chosen to be near the upper edge of the observed transition region from chaotic to periodic/quasi-periodic dynamics.
 Results for a total simulation time $t=2\times10^6$ and $t_{\rm trans}=10^6$.}
 \label{fig:21}
\end{figure}

This observation can be explained through a quantitative change in the systems 
dynamics in the presence of noise. As we have seen before, the dynamics changes from chaotic to periodic or quasi-periodic via an intermittent 
behavior.  For $N=10$, intermittency occurs even without noise.
This is nicely illustrated by the $x_N(t)$ graphs for different $u$ values in Fig. \ref{fig:20}. 
The intermittency found in the transition interval $27< u< 40$ explains why the 
sharp phase transition-like behavior disappears for larger systems. As the disorder parameter measures the fluctuations of the chain 
length, in the intermittent region smaller disorder parameters will be observed. Fig. 
\ref{fig:20} indicates that larger and larger periodic or quasi-periodic time intervals appear at increased 
chain velocities. 
Therefore, the sharp transition becomes smoother. For increasing system sizes, the previously 
observed noise induced phase transition-like behavior becomes also less evident, as illustrated by 
Fig. \ref{fig:21}, for driving velocities above the transition region.  

\begin{figure*}
\includegraphics[width=8.3 cm,keepaspectratio=true]{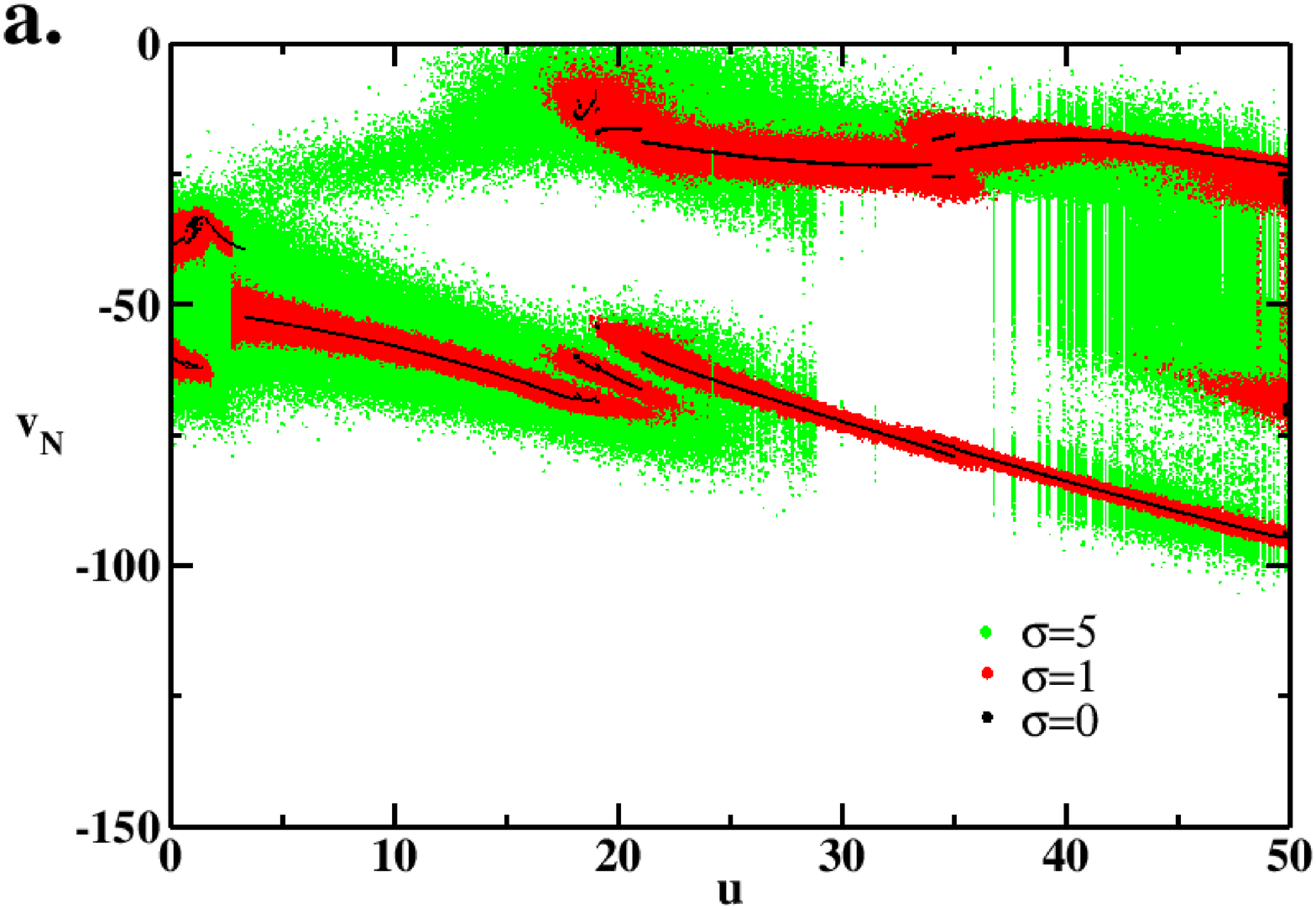}
\includegraphics[width=8.3 cm,keepaspectratio=true]{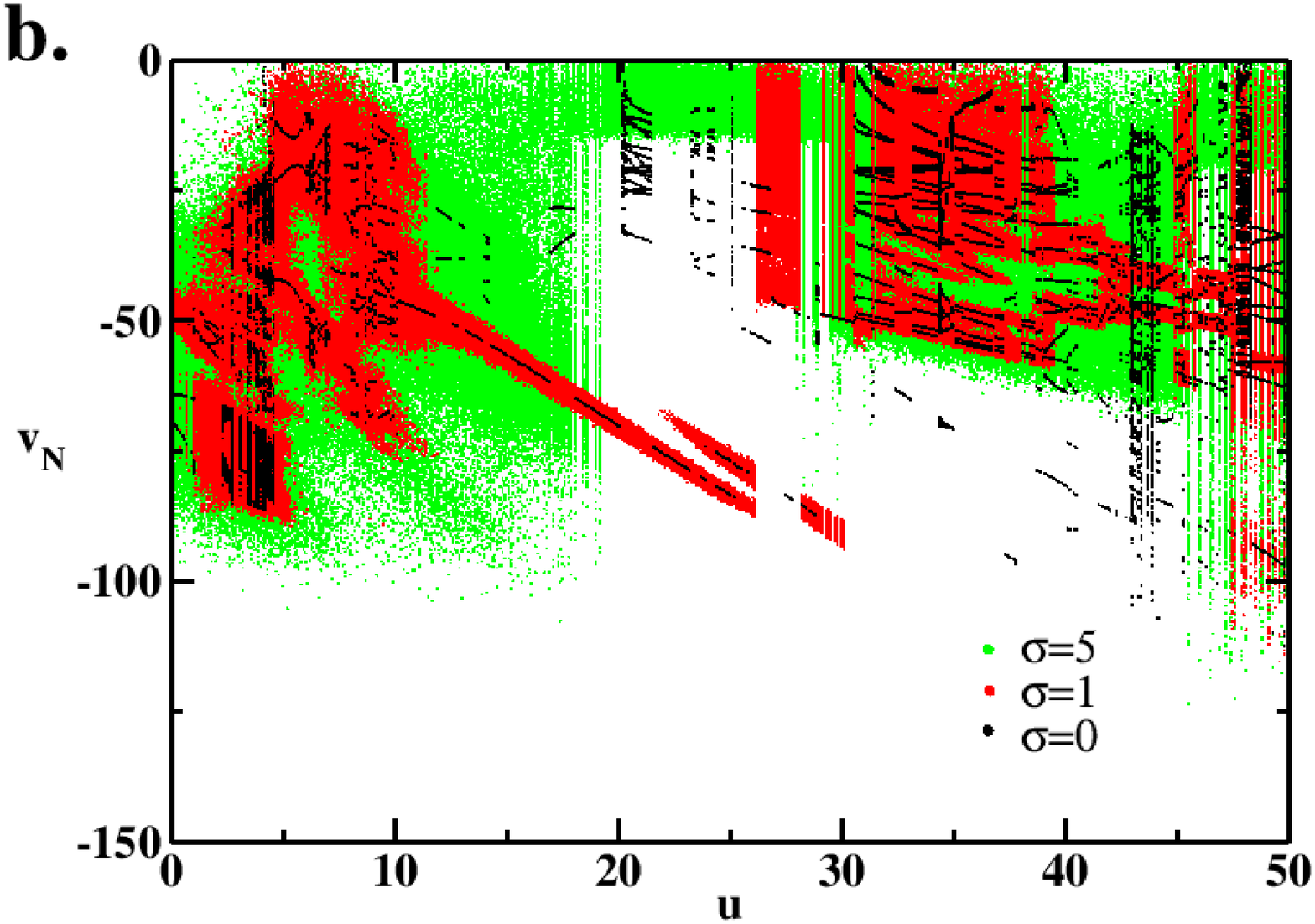}
\includegraphics[width=8.3 cm,keepaspectratio=true]{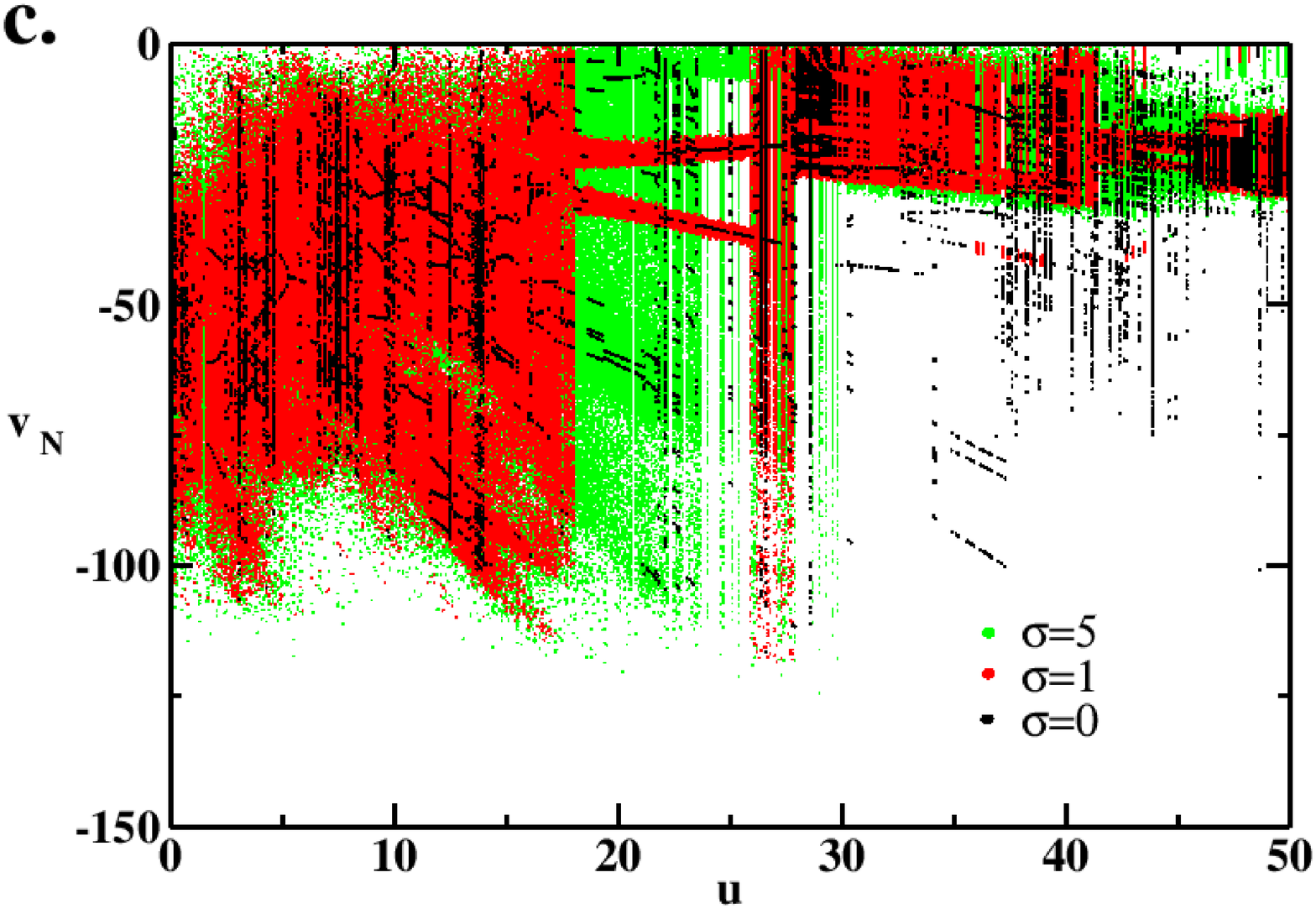}
\includegraphics[width=8.3 cm,keepaspectratio=true]{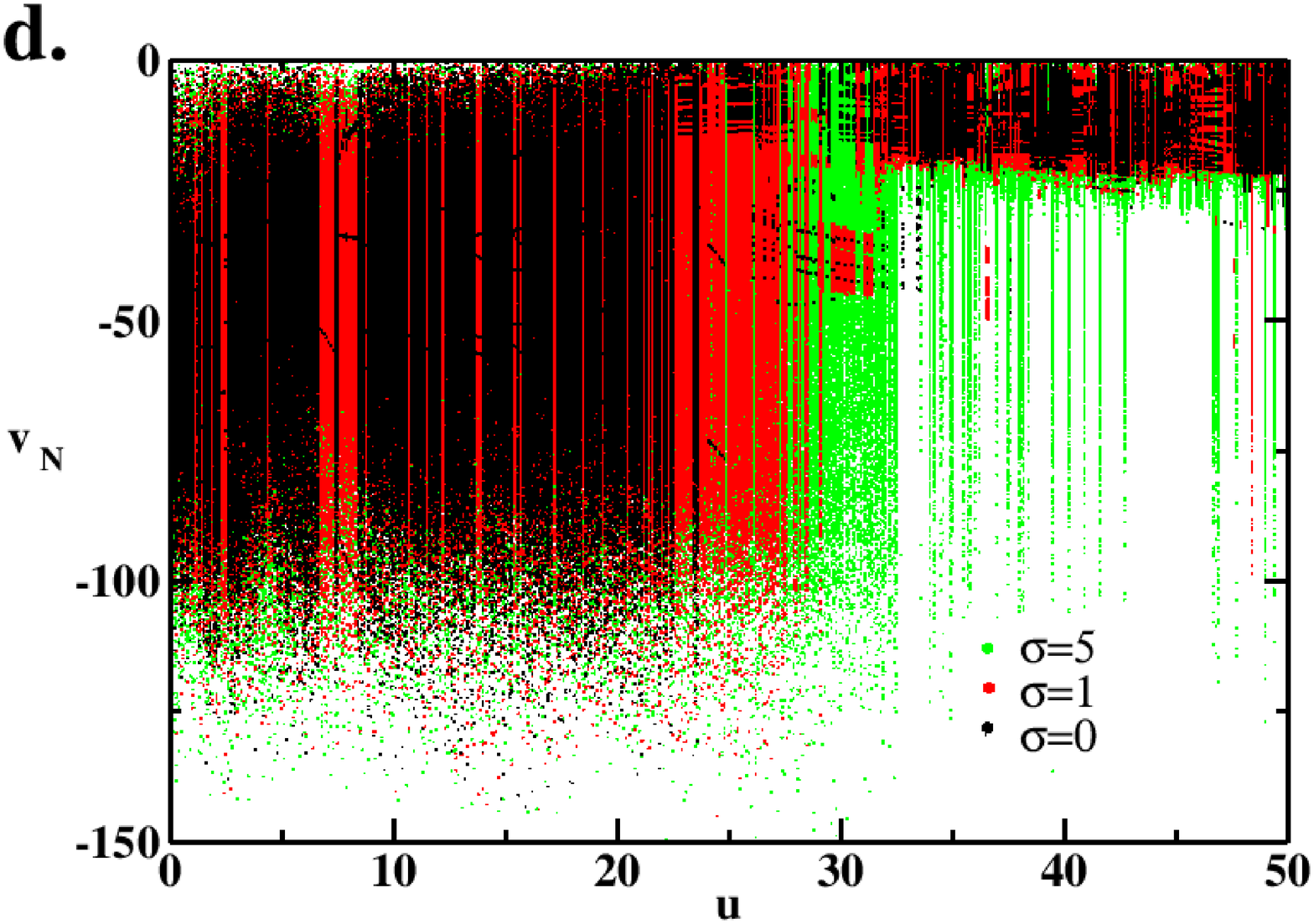}
\includegraphics[width=8.3 cm,keepaspectratio=true]{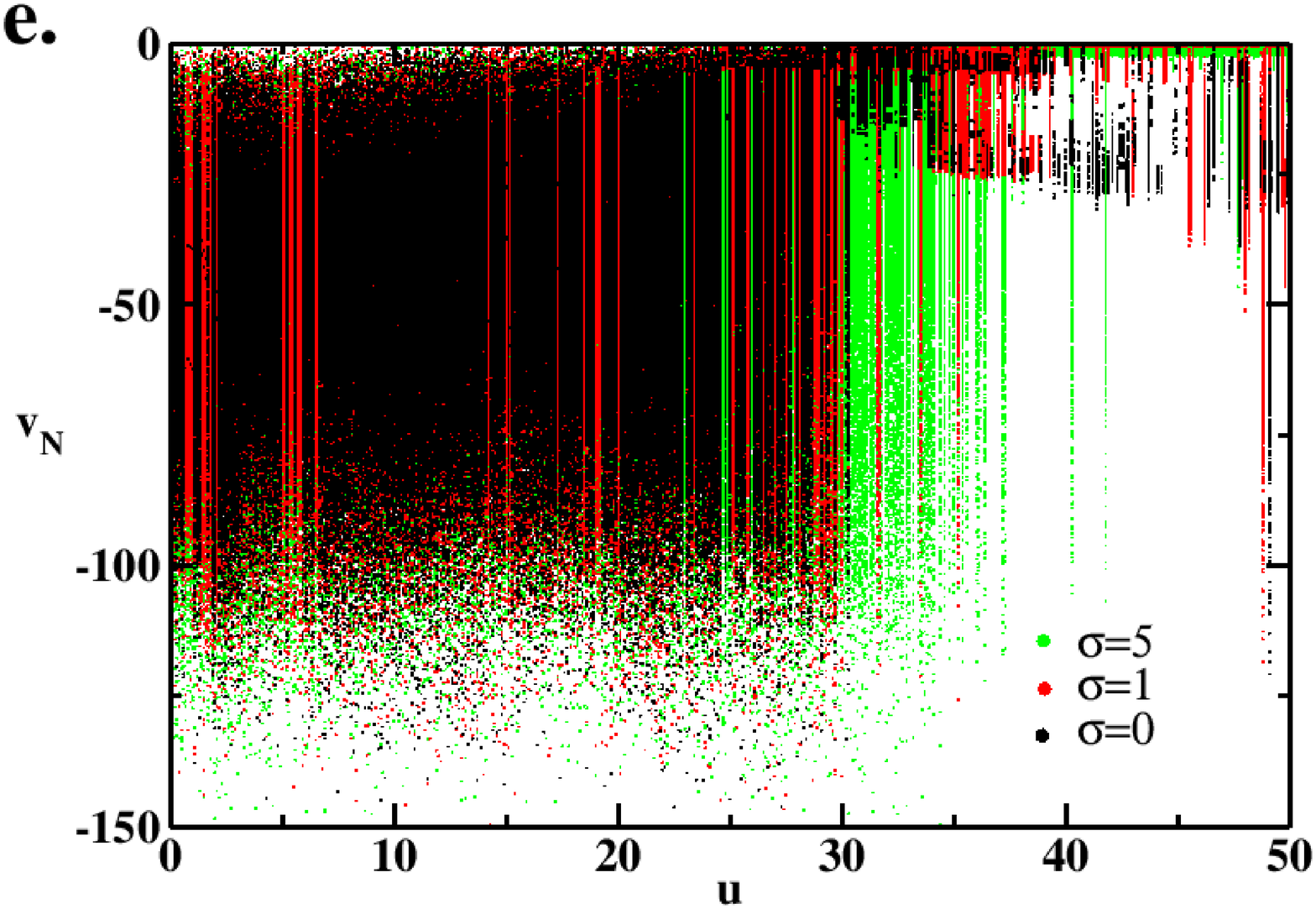}
\includegraphics[width=8.3 cm,keepaspectratio=true]{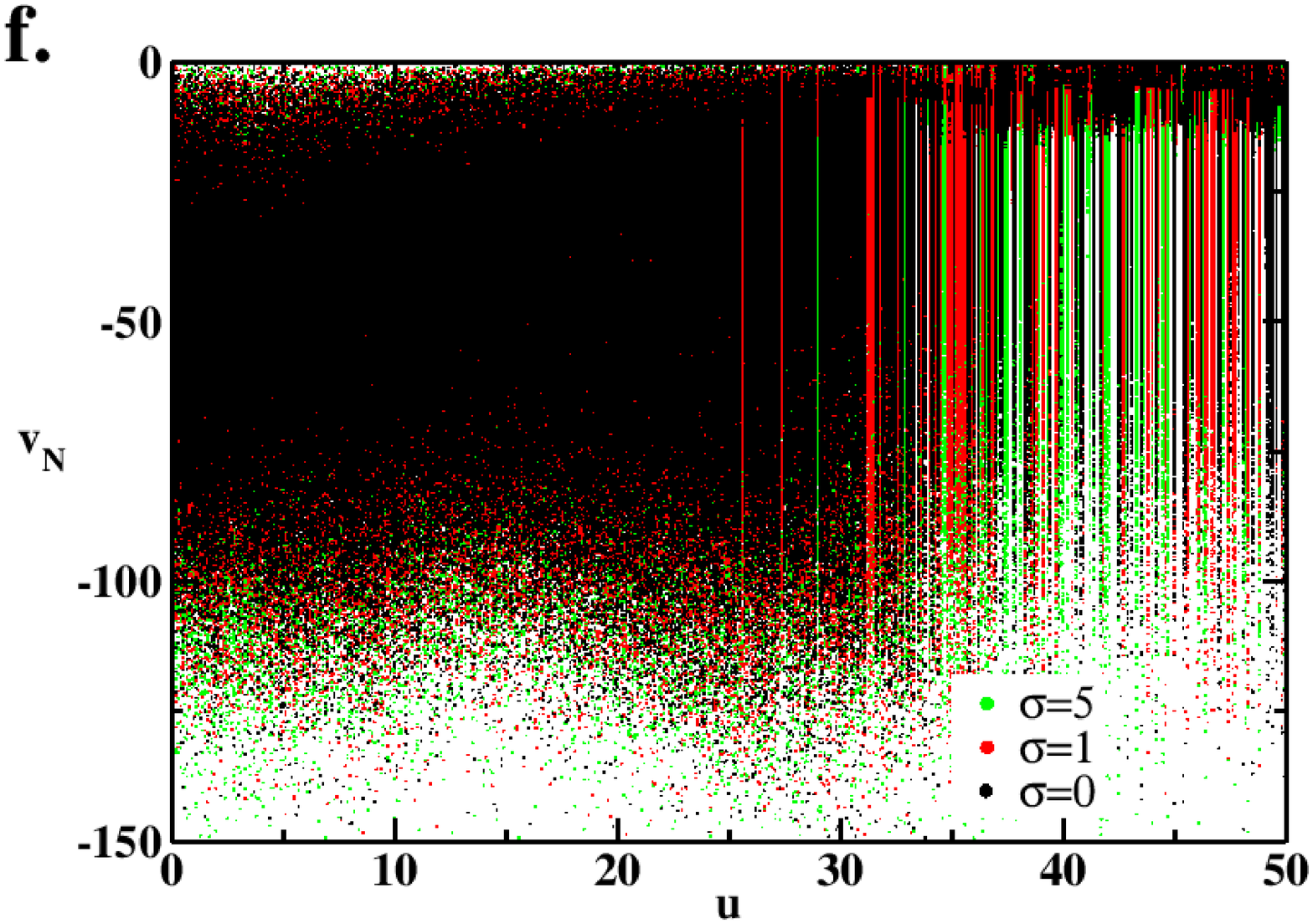}
 \caption{(color online) Bifurcation diagrams for different system sizes (a. $N=2$; b. $N=3$; c. $N=4$; d. $N=6$; e. $N=7$; f. $N=10$), 
and  different $\sigma$ values specified in the legends.}
 \label{fig:22}
\end{figure*}

Bifurcation diagrams generated for different system sizes are shown in Fig. \ref{fig:22}. 
These results augment those obtained from disorder parameter. They indicate that for $N<5$, 
the system has only periodic or quasi-periodic attractors. 
The behavior of the disorder parameter does not suggest any phase transition-like behavior here.
Dominant chaotic regimes appear for $N\geqslant 5$. 
As the size of the system is increased, the number of periodic windows decreases, and the transition region is enlarged.
Also, for larger systems the effect of noise is less pronounced. 

\section{Discussion}
\label{sec:conclusions}

The dynamics of a simple spring-block chain placed on a running conveyor belt was investigated both by simple experiments and 
through computer simulations. Despite its simplicity, the dynamics of the system proved to be quite complex exhibiting- chaotic, 
periodic or quasi-periodic behavior as a function of the conveyor belt's velocity and the amount of noise in the friction forces.
The experiments and computer simulations indicate that the transition from chaotic to periodic or quasi-periodic dynamics is typically realized through an 
intermittent chaotic state. 

In the chaotic regime, the  avalanche-size 
distribution function shows a scale-free nature, indicating also the presence of SOC, along with a $1/f$ type of stochasticity in the chain length.
Another aspect of the systems complexity is the diversity of the possible dynamical states reflected by the convoluted graph of the bifurcation diagram,
and a phase transition-like behavior in the disorder parameter. 
The presence of noise adds to this picture fascinating phenomena, like noise induced chaos, noise induced intermittency and noise induced phase transition.

Interestingly, the computer simulations suggest that the maximal complexity of the dynamical states is achieved for a relatively small 
number of blocks in the vicinity of $N=5$  for $\sigma<2$. For $N<5$ there is practically no chaos in the system.
With $N=5$ the observed transition is the sharpest one, and for $N>5$ it becomes smoother in a somewhat counter intuitive way.
It is remarkable that this collective behavior finds its explanation in terms of dynamical system properties.

Finally we would like to draw the attention to the fact that in several recent experimental studies \cite{Mudrock2011,Roth2012} performed on metallic alloys, it has been observed that just before the
onset of the plastic instability  the local strain rate as a function of the corresponding force response shows a multi periodic behavior. This is somehow similar with the graphs shown
in Fig. \ref{fig:11}. This observation  brings us back to the possibility of modeling the Portevin- Le Chatelier effect with a  spring-block system \cite{Lebyodkin1995}. Indeed, if the plastic deformation is macroscopically
uniform in the absence of plastic instability, one can assume that it is governed by only a small number of collective degrees of freedom, similarly with the dynamics 
of the spring-block chain considered here. In such view the studied model and the obtained results can gain interest in the field of material science as well.

\begin{acknowledgments}
The useful comments and advices of G\'abor Dr\'otos and G\'abor Csern\'ak are acknowledged.
The work of FJ-SZ and ZN is supported from the IDEAS research grant: PN-II-ID-PCE-2011-3-0348.
The work of BS was subsidized by "Collegium Talentum" of Hungary and by the Excellence Bursary of BBU. 
The research of BS and TT is conducted in the framework of T\'AMOP 4.2.4.A/1-11-1-2012-0001 ’National Excellence Program’ and OTKA NK100296. 
Financial support for this program is provided jointly by the Hungarian State, the European Union and the European Social Fund.
\end{acknowledgments}

\newpage

\appendix*
\section{Numerical method}
\label{sec:numerical_methods}
We briefly describe here the method used to integrate the Newton-equations (\ref{eq:newton}) and to handle the discontinuous stick-slip dynamics of 
the blocks. If a block is stuck to the conveyor belt, then it will move together with it at constant velocity $u$. Therefore, the position 
of the $i$th block relative to the ground is calculated with the simple 
\begin{equation}
 x_i(t+dt)=x_i(t)+u\cdot dt,
\end{equation}
equation. When a block is slipping relative to the belt, the basic Verlet method
\begin{equation}
 x_i(t+dt)=2x_i(t)-x_i(t-dt)+a_i(t){dt}^2+O({dt}^4)
\end{equation}
is used to update its position. As can be seen, this is a third order method, which can be extended also to the velocity space \cite{Lazar2010} as:
\begin{eqnarray}
 v_i(t+dt)=\frac{x_i(t)-x_i(t-dt)}{dt}+ \nonumber \\
 +\frac{1}{6}dt[11a_i(t)-2a_i(t-dt)] +O({dt}^3).
\end{eqnarray}
The instance when the $i$th block sticks to the belt is found when the relative velocity $v_{r_i}$ changes its sign, while the instance 
when the block starts to slip is defined by the sign change of $F_{ex}-F_{st}$. A more complicated stochastic numerical method was 
also developed to handle the stick-slip dynamics, but was found not to significantly alter the presented results.

\bibliographystyle{apsrev4-1}

\providecommand{\noopsort}[1]{}\providecommand{\singleletter}[1]{#1}%

\end{document}